\documentclass[apj]{emulateapj}

\usepackage{graphicx}
\usepackage{t1enc}
\usepackage[varg]{txfonts}
\usepackage{epsfig}
\usepackage{rotating}
\usepackage{natbib}
\usepackage{color}

\newcommand{\Td}    {$T_\mathrm{d}$}

\newcommand{\Trot}  {$T_\mathrm{rot}$}
\newcommand{\Tkin}  {$T_\mathrm{kin}$}
\newcommand{\mum}   {$\mu$m}
\newcommand{\kms}   {km~s$^{-1}$}
\newcommand{\cmg}   {cm$^{2}$~g$^{-1}$}
\newcommand{\cmt}   {cm$^{-3}$}
\newcommand{\cmd}	{cm$^{-2}$}
\newcommand{\jpb}   {$\rm Jy~beam^{-1}$}    
\newcommand{\lo}    {$L_{\sun}$}
\newcommand{\lbol} {$L_{\mathrm{bol}}$}
\newcommand{\mo}    {$M_{\sun}$}
\newcommand{\hh}	{H$_2$}

\newcommand{\nh}    {NH$_3$}

\newcommand{\chtoh} {CH$_3$OH}
\newcommand{\hho} {H$_2$O}

\newcommand{\et}    {et al.}
\newcommand{\eg}    {e.\,g.,}
\newcommand{\ie}    {i.\,e.,}
\newcommand{\hii}   {H{\small II}}

\newcommand{\supa}  {$^\mathrm{a}$}
\newcommand{\supb}  {$^\mathrm{b}$}
\newcommand{\supc}  {$^\mathrm{c}$}
\newcommand{\supd}  {$^\mathrm{d}$}
\newcommand{\supe}  {$^\mathrm{e}$}
\newcommand{\supf}  {$^\mathrm{f}$}

\newcommand{\phnn}  {\phantom{0}\phantom{0}}

\newcommand{\phsy}   {\phantom{$-$}}

\definecolor{RED}{rgb}{1.0,0.0,0.0}

\shorttitle{Fragmentation in G14.2}
\shortauthors{Busquet et al.}

\begin{document}

\title{What is controlling the fragmentation in the Infrared Dark Cloud G14.225--0.506? \\ Different level of fragmentation in twin hubs}

\author{Gemma Busquet\altaffilmark{1,2,3}$^\dagger$, Robert Estalella\altaffilmark{3}$^\dagger$, Aina Palau\altaffilmark{4}, Hauyu Baobab Liu\altaffilmark{5,6}, Qizhou Zhang\altaffilmark{7}, Josep Miquel Girart\altaffilmark{1,7}$^\dagger$, Itziar de Gregorio-Monsalvo\altaffilmark{6,8}, Thushara Pillai\altaffilmark{9}, Guillem Anglada\altaffilmark{2}, Paul T.~P. Ho\altaffilmark{5,7}}

\altaffiltext{1}{Institut de Ci\`encies de l'Espai (CSIC-IEEC), Campus UAB, Carrer de Can Magrans, S/N 08193 , Cerdanyola del Vallès, Catalunya, Spain}
\email{busquet@ice.cat}
\altaffiltext{2}{Instituto de Astrof\'isica de Andaluc\'ia, CSIC, Glorieta de la Astronom\'ia, s/n, E-18008, Granada, Spain}
\altaffiltext{3}{Departament d'Astronomia i Meteorologia, Institut de Ci\`encies del Cosmos (ICC), Universitat de Barcelona (IEEC-UB), Mart\'i i Franqu\`es, 1,
E-08028 Barcelona, Catalunya, Spain}
\altaffiltext{4}{Instituto de Radioastronom\'ia y Astrof\'isica, Universidad Nacional Aut\'onoma de M\'exico, P.O. Box 3-72, 58090 Morelia, Michoac\'an, M\'exico}
\altaffiltext{5}{Academia Sinica Institute of Astronomy and Astrophysics, Taipei, Taiwan}
\altaffiltext{6}{European Southern Observatory (ESO), Karl-Schwarzschild-Str. 2, D-85748 Garching, Germany} 
\altaffiltext{7}{Harvard-Smithsonian Center for Astrophysics, 60 Garden Street, Cambridge, MA 02138, USA}
\altaffiltext{8}{Joint ALMA Observatory (JAO), Alonso de C\'ordova 3107, Vitacura, Santiago de Chile, Chile}
\altaffiltext{9}{Max Planck Institut f\"ur Radioastronomie, Auf dem H\"ugel 69, D-53121 Bonn, Germany}
\altaffiltext{$^\dagger$}{The ICCUB is a CSIC-Associated Unit through the ICE}

\begin{abstract}
We present observations of the 1.3~mm continuum emission toward hub-N and hub-S of the infrared dark cloud G14.225--0.506 carried out with the Submillimeter Array, together with observations of the dust emission at 870~\mum\ and 350~\mum\ obtained with APEX and CSO telescopes. 
The large scale dust emission of both hubs consists of a single peaked clump elongated in the direction of the associated filament. At small scales, the SMA images reveal that both hubs fragment into several dust condensations. The fragmentation level was assessed under the same conditions and we found that hub-N presents 4 fragments while hub-S is more fragmented, with 13 fragments identified. 
We studied the density structure by means of a simultaneous fit of the radial intensity profile at 870 and 350~\mum\ and the spectral energy distribution adopting a Plummer-like function to describe the density structure. The parameters inferred from the model are remarkably similar in both hubs, suggesting that density structure could not be responsible in determining the fragmentation level. We estimated several physical parameters such as the level of turbulence and the magnetic field strength, and we found no significant differences between these hubs. The Jeans analysis indicates that the observed fragmentation is more consistent with thermal Jeans fragmentation compared with a scenario that turbulent support is included. The lower fragmentation level observed in hub-N could be explained in terms of stronger UV radiation effects from a nearby \hii\ region, evolutionary effects, and/or stronger magnetic fields at small scales, a scenario that should be further investigated. 
\end{abstract}

\keywords{ISM: clouds -- stars: formation -- 
ISM: individual objects (G14.225--0.506)
}

\section{Introduction \label{sint}}

One ubiquitous fact in the process of formation of intermediate/high-mass stars is that they are usually found 
associated with clusters of lower-mass stars
\citep[\eg][]{pudritz2002,lada2003}. However, the process behind the
formation of  such clusters is unclear as it remains ambiguous how a cloud core fragments to finally form a cluster. 
A common approach is to assume that the fragmentation is controlled by gravitational instability, where the velocity dispersion 
is, at best, accounted for only as an extra source of pressure support. 
In massive star-forming regions several studies show that thermal fragmentation 
alone does not account for the observed masses and/or core (fragment) separation
\citep[\eg][]{zhang2009,bontemps2010,pillai2011,wang2011,naranjo-romero2012,vanKempen2012,wang2014,zhang2015,liu2015}. Most of these works only address the formation of massive dense cores, which are far more massive than the thermal Jeans mass, and thus do not follow thermal Jeans fragmentation. However, in stellar clusters, most of stars are low-mass stars and their masses are related to thermal Jeans mass, as found by \citet{takahashi2013},  \citet{palau2015}, and \citet{teixeira2015},
raising  the question of \textit{what is controlling the fragmentation
  process?} The observational results suggest that cloud fragmentation
leading to the formation of pre-stellar cores must be controlled by a complex interaction of gravitational instability, turbulence, magnetic 
fields, cloud rotation, and stellar feedback \citep[\eg][]{padoan2002,hosking2004,machida2005,girart2013}.

The infrared dark cloud (IRDC) G14.225--0.506 (hereafter G14.2), also known as M17 SWex \citep{povich2010}, is part of an extended (77\,pc$\times16$\,pc) and massive ($>$10$^5$\,\mo) 
molecular cloud discovered by \citet{elmegreen1976} and located
southwest of the Galactic \hii\ region M17. The distance to M17 has been recently
determined through trigonometric parallaxes of \chtoh\ masers, to be 1.98$^{+0.14}_{-0.12}$\,kpc \citep{xu2011,wu2014}.
Similar local standard of rest velocity  \citep{elmegreen1979,busquet2013} suggests that the bright \hii\ region M17 and the IRDC are located at the same distance. 
The cloud is associated with star formation activity as revealed by the presence of several \hho\ masers \citep{jaffe1981,palagi1993,wang2006} and 
intermediate-mass 
YSOs \citep{povich2010}. Through the analysis of near-/mid-infrared
photometry \textit{Spitzer} data, \citet{povich2010} report an absence
of early O stars and suggest a delay in the onset of massive star formation in the region, concluding 
that G14.2 represents a proto-OB association that has not yet formed its most 
massive stars. 

\begin{figure}[!t]
\begin{center}
\begin{tabular}[t]{c} 
	\epsfig{file=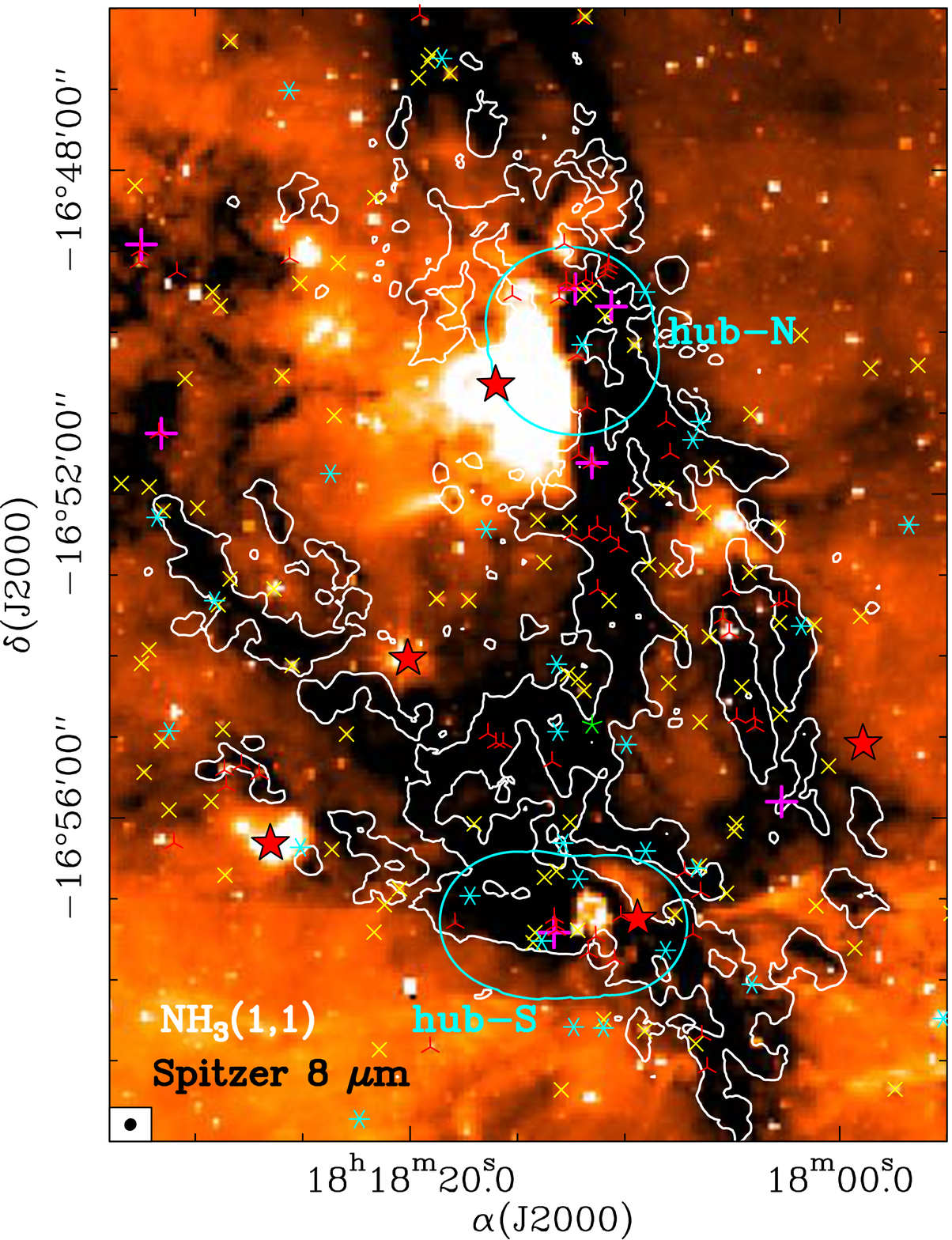,scale=0.44,angle=0}	\\
	\end{tabular}
\caption{Overview of the large scale structure of the G14.2 complex. White contours represent the $3\sigma$ contour level (27~m\jpb\,\kms) of the \nh\,(1,1) integrated
intensity map \citep{busquet2013}. The \nh\ synthesized beam, $8\farcs0\times7\farcs2$ (P.~A.$=-15\degr$) is shown in the bottom left corner. 
Color scale is the
8~\mum\ \textit{Spitzer} image. Red stars indicate
IRAS sources in the field, and pink crosses mark the position of \hho\ masers \citep{wang2006}. 
Color stars depict the positions of the YSOs identified 
by \citet{povich2010} where colors and symbols indicate the YSO evolutionary stage; three-point red star: Stage 0/I (dominated by infalling envelope); 
four-point yellow star: Stage II (optically thick
circumstellar disk); five-point green star: Stage III (optically thin disk); six-point cyan star: ambiguous (see \citealt{povich2009,povich2010} for further details on the YSOs 
classification). The SMA field of view of the two regions mosaiced, hub-N and hub-S, are indicated with the blue contour.}
\label{g14-nh311}
\end{center}
\end{figure}

Recent high angular resolution observations of the \nh\ 
dense gas, obtained by combining VLA and Effelsberg~100\,m data \citep{busquet2013}, reveal a network of filaments constituting two hub-filament 
systems. 
Hubs are associated with \hho\ maser emission \citep{wang2006} and
mid-infrared sources \citep{povich2010}. They appear more compact, are
warmer (\Trot$\sim$15~K) and show larger velocity dispersion and larger masses per unit length than filaments \citep{busquet2013}, 
suggesting that they are the main sites of stellar activity within the cloud.
In Figure~\ref{g14-nh311} we provide a general perspective of the cloud, showing the dense filaments and hubs and their association with IR sources and/or water masers. 

In this paper we present 1.3~mm observations performed with the 
Submillimeter Array (SMA\footnote{The Submillimeter Array is a joint project between the Smithsonian Astrophysical Observatory 
and the Academia Sinica Institute of Astronomy and Astrophysics, and is funded by the Smithsonian Institution and the Academia Sinica.}), 870~\mum\ observations carried out with the LABOCA bolometer at the APEX telescope, and 350~\mum\ observations with the SHARC~II bolometer at the Caltech Submillimeter Observatory (CSO), 
toward the two hubs, dubbed hub-N and hub-S (see Fig.~\ref{g14-nh311}). These observations are aimed at studying the cloud fragmentation process that 
leads to the formation of dense core fragments.
Section~2 describes the observations and 
data reduction process. In Section~3 we present the results of the dust continuum emission observations, and their analysis is presented in Section~4. 
We discuss the interplay between density, turbulence, magnetic fields, UV radiation feedback, and evolutionary effects in the fragmentation process in Section~5. 
Finally, in Section~6 we list the main conclusions.

\begin{table*}[!ht]
\begin{footnotesize}
\caption{SMA observational parameters}
\centering
\begin{tabular}{l c c c c c c c c c c}
\hline\hline\noalign{\smallskip}	

Source &Date	&Configuration &Number &Proj. baselines  
&Gain   &Bandpass &Flux &$\nu_{\mathrm{LO}}$\supa &Continuum bandwidth  \\ 
&&&of Antennas &(m)  & calibrator
& &calibrator & (GHz) &(GHz) \\
\hline
\noalign{\smallskip}
Hub-S	&2008/06/29 	&compact	 		&8 	&16--139			
&1733$-$130/1911$-$201		&3c273/3c454.3		&Uranus &221.98 	&3.6	\\ 
Hub-S	&2008/07/21	&extended 		&7	&44--226			
&1733$-$130/1911$-$201		&3c273/3c454.3		&Uranus &225.42    &3.4	 \\ 
Hub-N	&2011/04/19	&compact 		&8 	&16--\phn77 		
&1733$-$130/1924$-$292 		&3c454.3				&Neptune	 &224.66	&7.8 \\	
Hub-N\supb\	&2015/04/04		&extended		&7	&23--226  &1733$-$130	&3c273	&Titan	&225.42	&7.6	\\
Hub-N	&2014/08/21	&very extended 	&5 	&68--432			
&1733$-$130					&3c279				&Ceres	 &225.47	&8.0	\\	
\hline
\end{tabular}
\begin{list}{}{}
\item[\supa]{Center local oscilator frequency.}
\item[\supb]{Data taken with the new correlator SWARM, although only data belonging to the old ACIS correlator are used in this work. }
\end{list}
\label{tsmaobs}
\end{footnotesize}
\end{table*}

\begin{table}[!ht]
\caption{SMA imaging parameters}
\centering
\begin{tabular}{l c c c c}
\hline\hline\noalign{\smallskip}	
Source &Configuration 	&$\theta_{\mathrm{LAS}}$\supa	&Synthesized beam  	& Rms noise\supb  \\
&&			          							&FWHM, P.~A\supc 	&	\\	
&&					(arcsec)					          &(arcsec, $\degr$)	&(m\jpb)	\\		
\hline
\noalign{\smallskip}
Hub-S	&compact	 		&7	&$3\farcs09\times2\farcs28$, \phsy57$\degr$        	&2.8 	 \\	
Hub-S	&extended 		&3	&$1\farcs23\times1\farcs09$, $-5\degr$	 	&1.2 	 \\	
Hub-S	&combined		&7	&$1\farcs53\times1\farcs41$,  \phsy35$\degr$	&1.0 	  \\
Hub-N	&compact 		&7	&$4\farcs13\times2\farcs75$, \phsy35$\degr$		&2.0	 \\	
Hub-N	&extended		&4	&$1\farcs40\times0\farcs97$, $-$62$\degr$		&2.0 	 \\	
Hub-N	&very extended 	&2	&$1\farcs57\times0\farcs46$, $-79\degr$			&3.5		 \\
Hub-N	&combined		&7	&$1\farcs46\times0\farcs46$, $-78\degr$		 &1.0\\
\hline
\end{tabular}
\begin{list}{}{}
\item[\supa]{Largest angular structure to which an interferometer is sensitive $[\theta_{\mathrm{LAS}}/\mathrm{arcsec}]$=91.02$[u_{\rm min}/k\lambda]^{-1}$, where $u_{\rm min}$ is the shortest baseline \citep{palau2010}.}
\item[\supb]{Root-mean-square noise level of the continuum data.}
\item[\supc]{FWHM: full-width at half-maximum; P.~A.: position angle.}
\end{list}
\label{tsmamaps}
\end{table}


\section{Observations and Data Reduction \label{sobs}}

\subsection{Submillimeter Array Observations}

The observations were carried out with the SMA \citep{ho2004}  between 2008 and 2015 
in different array configurations. 
Table~\ref{tsmaobs} lists the observation dates, the configuration used, the number of antennas, and the projected baselines for each observing date. 
We performed a mosaic of 3 pointings toward hub-S and of 2 pointings toward hub-N,
covering an area of $\sim2'\times1'$ and $\sim1'\times1\farcm5$, respectively (see blue contour in Fig.~\ref{g14-nh311}).

For hub-S, the SMA correlator covered a 2~GHz bandwidth in each of the two sidebands, which are separated by 10~GHz. Each sideband is divided into 6 blocks, 
and each block consists of 4 chunks, yielding 24 chunks of 104~MHz width per sideband.  
For both compact and extended configurations the correlator was set to the standard mode with 256 channels
per chunk, providing uniform channel spacing of 0.406~MHz (or 0.52~\kms) per channel across the full bandwidth of 2~GHz. 
On the other hand, for hub-N the observations were conducted as filler projects in different runs 
using the 4~GHz receivers. In this case, each sideband 
was divided into 12 blocks with 4 chunks of 104~MHz each.
The configuration of the correlator was different for each observing run. Table~\ref{tsmaobs} summarizes the details of the SMA correlator, listing the local oscillator (LO) center frequency and the total bandwidth used to build the continuum.  

The zenith opacity ($\tau_{\rm{225\,GHz}}$), measured with the National Radio Astronomy Observatory (NRAO) tipping radiometer located at the Caltech 
Submillimeter Observatory, was stable during the observations in the compact configurations, with values $\sim0.1$-0.2, and around 0.1 during the observations in 
the extended and very extended configurations. 

The visibility data were calibrated using the standard procedures of MIRIAD \citep{sault1995}. Information on the calibrators used is given in Table~\ref{tsmaobs}. The flux calibration is estimated to be accurate within $\sim20$\,\%.

The continuum was constructed from the line-free channels in 
the visibility domain. To produce the CLEANed single images of the two hubs, we used the tasks ``invert'' (option 
mosaic), ``mossdi'', and ``restor'' as recommended in the MIRIAD users guide\footnote{http://www.atnf.csiro.au/computing/software/miriad/userguide/userhtml.html} 
\citep{sault1995}. The final combined images were obtained by weighting the visibilities in inverse proportion to the noise variance.
We summarize the basic parameters (synthesized beam and rms noise level) of the different maps in Table~\ref{tsmamaps}.
In this work we focus on the dust continuum 
emission, leaving the analysis of the molecular line emission for a forthcoming paper.

\subsection{APEX Observations}

Continuum observations at 870~\mum\ were carried out using LABOCA bolometer
array, installed on the Atacama Pathfinder EXperiment (APEX\footnote{This work is partially based on observations with the APEX telescope. 
APEX is a collaboration between the Max-Plank-Institute f\"ur Radioastronomie, the European Southern Observatory, and
the Onsala Space Observatory.}) telescope. The
array consists of 259 channels, which are arranged in 9 concentric hexagons around
the central channel. The field of view of the array is $11\farcm4$, and the angular
resolution of each beam is $18\farcs6\pm1''$.

The data were acquired on 2008 August 24 and 31 during the ESO program
081.C-0880A, under excellent weather conditions (zenith opacity values ranged from
0.15 to 0.24 at 870~\mum). Observations were performed using a spiral raster mapping.
This observing mode consists of a set of spirals with radii between $2'$ and $3'$ at a
combination of 9 and 4 raster positions separated by $60''$ in azimuth and elevation,
with an integration time of 40 seconds per spiral. This mode provides a fully sampled
and homogeneously covered map in an area of $15'\times15'$. The final map consisted of a
mosaic of two points centered at $\alpha$(J2000)=18$^{\mathrm{h}}$18$^{\mathrm{m}}$17$^{\mbox{\scriptsize{s}}}$.5, $\delta$(J2000)=$−16\degr44'00\farcs0$,
and $\alpha$(J2000)=18$^{\mathrm{h}}$18$^{\mathrm{m}}$17$^{\mbox{\scriptsize{s}}}$.5, $\delta$(J2000)=$−16\degr57'00\farcs0$, respectively.
The total on-source integration time was $\sim2$~hours per position. Calibration was performed using
observations of Mars as well as secondary calibrators. The absolute flux calibration
uncertainty is estimated to be $\sim8$~\%. The telescope pointing was checked every
hour, finding a rms pointing accuracy of 2$''$. Focus settings were checked once per
night and during the sunset.

We reduced the data using MiniCRUSH software package \citep[see][]{Kov08}. The pre-processing steps consisted of flagging dead or cross-talk channels frames
with too high telescope accelerations and with unsuitable mapping speed, as well
as temperature drift correction using two blind bolometers. Data reduction process
included flat-fielding, opacity correction, calibration, correlated noise removal (atmospheric
fluctuations seen by the whole array, as well as electronic noise originated
in groups of detector channels), and de-spiking. Every scan was visually inspected to
identify and discard corrupted data. The final map was smoothed to a final angular
resolution of $\sim22''$, and the rms noise level achieved was 25~m\jpb. 
In this paper we focus our attention on the dust emission associated with the two hubs, hub-N and hub-S, identified using high angular resolution \nh\ data
\citep{busquet2013}.
The analysis of the entire cloud will be the subject of a forthcoming
paper.

 \begin{table*}[ht]
\caption{Parameters of hubs from APEX-LABOCA 870~\mum\ and CSO-SHARC~II 350~\mum\ observations}
\centering
\begin{tabular}{l c c c c c c c}
\hline\hline\noalign{\smallskip}	
  	&$\alpha$(J2000.0)	&$\delta$(J2000.0) &I$_{\nu}^{\mathrm{peak}}$ 		&S$_{\nu}$ 	&\multicolumn{2}{c}{Deconvolved Size}  &P.A.    \\
\cline{6-7}
 Source		&(h:m:s) 	&($\degr$:$'$:$''$)  &(\jpb) 				&(Jy)		&(arcsec)			&(pc) 	  &(\degr)  \\
\hline
\noalign{\smallskip}
870~\mum\ \\ 
\hline
hub-N	&18:18:12.62 &$-$16:49:34.0 	&\phn6.4$\pm0.1$    			&\phn24.7$\pm$2.2  		&57.7$\times$33.2		&0.6$\times$0.3	&\phnn7.2	 \\
hub-S	&18:18:13.10 &$-$16:57:20.3	&\phn4.5$\pm0.1$			        &\phn20.0$\pm$1.9  		&63.0$\times$33.4		&0.6$\times$0.3 	&\phn81.5	\\
\hline
\noalign{\smallskip}
350~\mum\ \\
\hline
hub-N	&18:18:12.53	&$-$16:49:28.2	 &17.0$\pm0.2$   				&111.6$\pm$7.7			&30.3$\times$20.7		&0.3$\times$0.2	&164.4 \\
hub-S	&18:18:13.06	&$-$16:57:19.2  &10.1$\pm0.2$				&\phn81.3$\pm$7.5			&29.9$\times$20.9 		&0.3$\times$0.2	&\phn71.8	\\
\hline
\end{tabular}
\label{tclumpprop}
\end{table*}

\subsection{CSO SHARC~II 350~\mum} \label{subsec:cso}

High angular resolution continuum observations at 350~\mum\ were carried out using the SHARC~II bolometer array, installed on the Caltech Submillimeter Observatory (CSO\footnote{This work is based upon work at the Caltech Submillimeter Observatory, which is operated by the California Institute of Technology.}).
The array consists of 12$\times$32 pixels (approximately 85~\% of these pixels work well).
The simultaneous field of view provided by this array is $2\farcm59\times0\farcm97$, and the diffraction limited beam size is $\sim9''$.

The data were acquired on 2014 March 26 ($\tau_{\mbox{\scriptsize{225~GHz}}}$$\simeq$0.08).
The telescope pointing and focusing were checked every 1.5--2.5~hours. Mars was observed for the absolute flux calibration, with a flux calibration uncertainty of $\sim20$~\%. We used the standard 10$'\times10'$ on-the-fly (OTF) box scanning pattern, centered on the two positions $\alpha$(J2000)=18$^{\mbox{\scriptsize{h}}}$18$^{\mbox{\scriptsize{m}}}$13$^{\mbox{\scriptsize{s}}}$.99; $\delta$(J2000)=$-16^{\circ}$51$'$00$\farcs$40, and $\alpha$(J2000)=18$^{\mbox{\scriptsize{h}}}$18$^{\mbox{\scriptsize{m}}}$13$^{\mbox{\scriptsize{s}}}$.997; $\delta$(J2000)=$-16^{\circ}$59$'$00$\farcs$40. The final map covered an area of $20'\times10'$. 
The total on-source time was 30 minutes for each of these two pointings. Data calibration was performed using the CRUSH software package (Kov\'{a}cs 2008). The final map was smoothed to an angular resolution of $9\farcs$6 and the rms noise level achieved was $\sim85$~mJy\,beam$^{-1}$.

\section{Results \label{sres}}

\subsection{Dust emission at 0.1~pc scale\label{largescale}}

\begin{figure*}[!ht]
\begin{center}
\begin{tabular}[t]{c} 
\epsfig{file=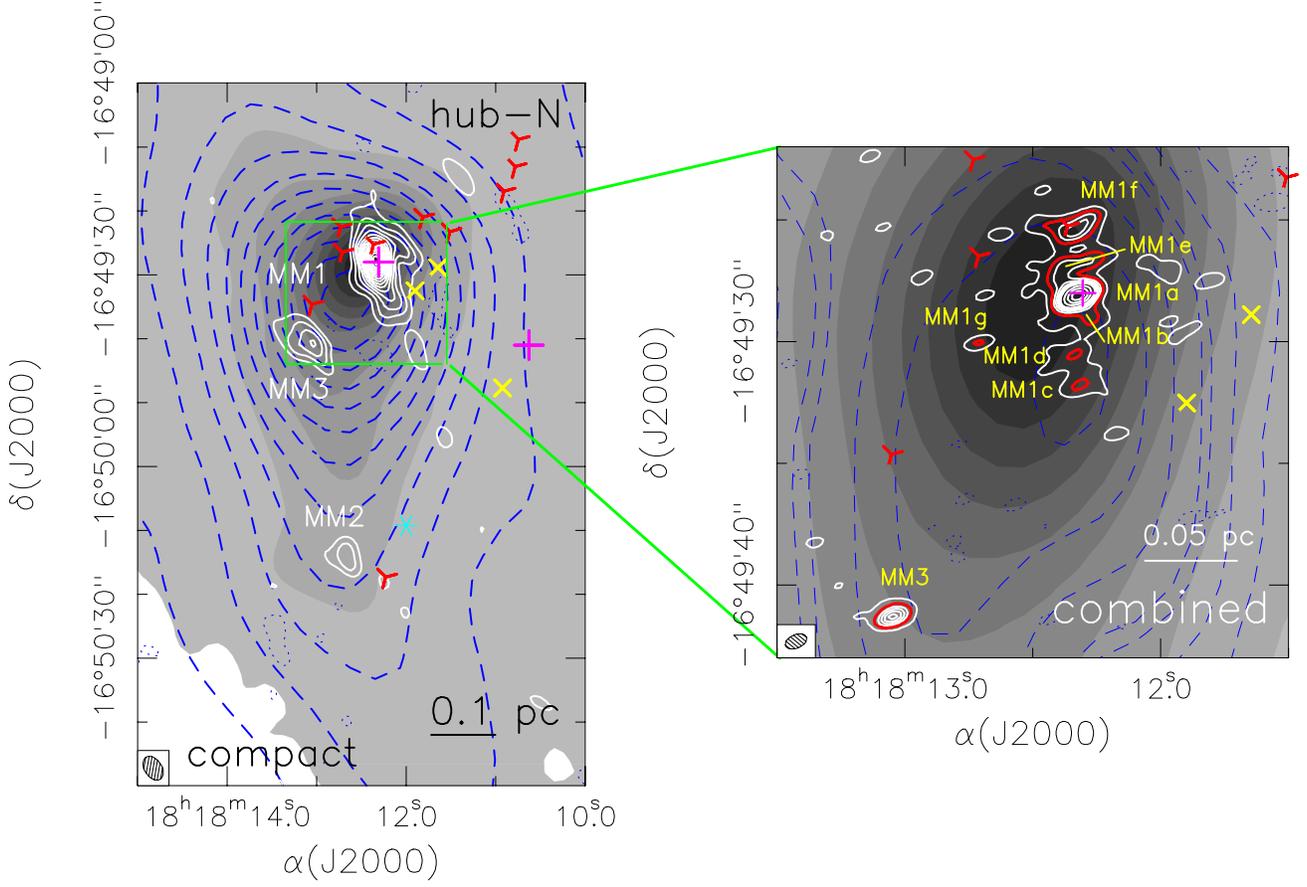,scale=0.85,angle=0} \\
	\end{tabular}
\caption{\emph{Left:} Contour map of the dust emission at 1.3~mm obtained with the SMA compact configuration (white contours, angular resolution $\sim$4$''$) of G14.2-hub-N 
overlaid on the dust emission at 350~\mum\ from the CSO SHARC~II bolometer (grey scale) and at  870~\mum\ from LABOCA bolometer at the APEX telescope (dashed contours). 
Contours at 1.3~mm start at $3\sigma$ and increase in steps of $3\sigma$, where $\sigma$ is the rms of the map listed in Table~\ref{tsmaobs}. Blue dotted contour corresponds to the $-3\sigma$ contour level. The dashed contours at 870~\mum\ start at $3\sigma$ rms level and increase in steps of $20\sigma$, where $\sigma$ is 25~m\jpb. The CSO 350~\mum\ grey-scale image start at 2~\% of the maximum, with levels increasing in 10~\% of the maximum value.
The synthesized beam of the SMA image is shown in the bottom left corner of the image. 
\emph{Right:} Close-up of the MM1 and MM3 millimeter sources that show the SMA 1.3~mm continuum emission obtained by combining all configurations. The red thick contour level denotes the 6\,$\sigma$ level used to identify the continuum sources. 
Contours range from 3 to $21\sigma$ in steps of $3\sigma$, and from 21 to $51\sigma$ in steps of $10\sigma$, where $\sigma$ is the rms of the map listed in Table~\ref{tsmaobs}.
Dashed contours represent the \nh\,(1,1) integrated intensity emission \citep{busquet2013} and the grey scale is the 350~\mum\  image. 
The synthesized beam ($1\farcs43\times0\farcs80$, P.~A.$=-54\degr$) is shown in the bottom left corner of the image.
Symbols are the same as in Fig.~\ref{g14-nh311}. 
}
\label{fsmacont-hubN}
\end{center}
\end{figure*}

Figure~\ref{fsmacont-hubN} (left panel) and Fig.~\ref{fsmahubS-comb} (top panel) show the dust emission at 870~\mum\ (blue dashed contours) and 350~\mum\ (grey scale) obtained with the APEX and the CSO telescopes toward hub-N and hub-S, respectively. The smaller scale structure of the dust continuum emission follows the larger scale filaments seen in extinction, \ie\ hub-N appears elongated in the north-south direction. 
Similarly, the large scale dust continuum emission of hub-S appears elongated
along the east-west direction. 
The morphology of the dust emission resembles that of the dense gas traced by \nh\  \citep{busquet2013}. Table~\ref{tclumpprop} summarizes the main physical parameters of the dusty envelope of both hubs, listing their peak position, peak intensity, and flux density. The deconvolved size was obtained by fitting a 2-dimensional Gaussian function. 
The two hubs have the same size, $0.6\times0.3$~pc at 870~\mum\ and
$0.3\times0.2$~pc at 350~\mum. Note that the peak positions at 870 and
350~\mum\ are offset from each other by $\sim6''$ in hub-N and $\sim3''$ in hub-S. This is most likely due to pointing errors. Further details on the physical properties of hubs are presented in Section~\ref{model} where we model the radial intensity profile and the spectral energy distribution of each hub.

\subsection{Dust emission at 0.03~pc scale\label{smacontres}}

In this section we present first the results obtained with the SMA in the compact configuration with an angular resolution of $\sim3''-4''$, corresponding to $\sim0.03$~pc at the distance of the source. Then, in Sect.~\ref{smahigh} we show the images of each hub obtained by combining all configurations. 
This allows us to obtain high angular resolution maps ($\sim1\farcs5$) at the best possible sensitivity. From these maps, we used a $6\sigma$ threshold to identify the millimeter condensations in each hub, with $\sigma$ being the rms noise level and requesting that the 6$\sigma$ contour level is closed.

Figure~\ref{fsmacont-hubN} (left panel) and Fig.~\ref{fsmahubS-comb} (top panel) present the SMA 1.3~mm continuum emission (in white contours) obtained with the compact configuration toward hub-N and hub-S, respectively.
At a similar angular resolution and similar sensitivity, 2-3~m\jpb, the dust envelopes of both hubs appear fragmented into dust cores. 
We detected 3 and 7 compact continuum condensations above a $6\sigma$ contour level toward hub-N and hub-S, respectively, in the compact configuration. 

In hub-N, most of the 1.3~mm emission arise from a bright source, MM1, located at the peak of the single-dish submillimeter dust emission. This appears to be elongated along the north-south direction with a position angle of $7\degr$, similar to the large scale 870~\mum\ emission (see Table~\ref{tclumpprop}). 
There are two faint and more flattened millimeter condensations associated with hub-N: MM2 located about $46''$ south of MM1, and MM3 located $\sim12''$ south-east of MM1. All millimeter condensations in hub-N are deeply embedded within the dust envelope and the \nh\,(1,1) dense gas emission. 

The 1.3~mm SMA continuum emission of hub-S consists of 7 continuum condensations, 4 of them clustered around the peak position of the large scale envelope, forming a snake-shape structure of about $25''$ (Fig.~\ref{fsmahubS-comb}-top panel). From west to east the millimeter condensations are labeled as MM2, MM3, MM4, and MM5.  

\subsection{Dust emission at 0.01~pc scale\label{smahigh} }

\begin{figure*}[!ht]
\begin{center}
\begin{tabular}[t]{c} 
\epsfig{file=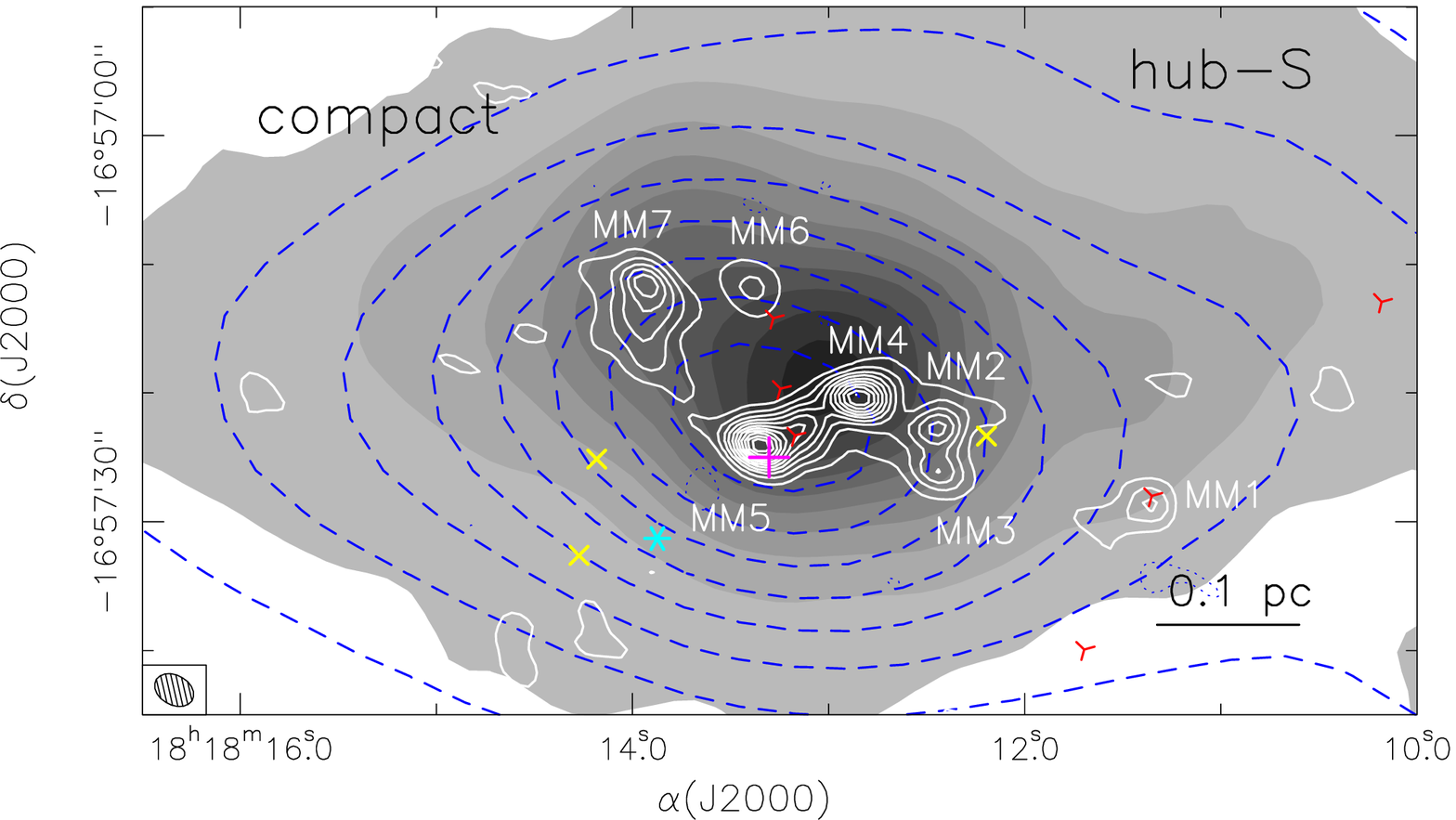,scale=0.7,angle=0}	\\
\epsfig{file=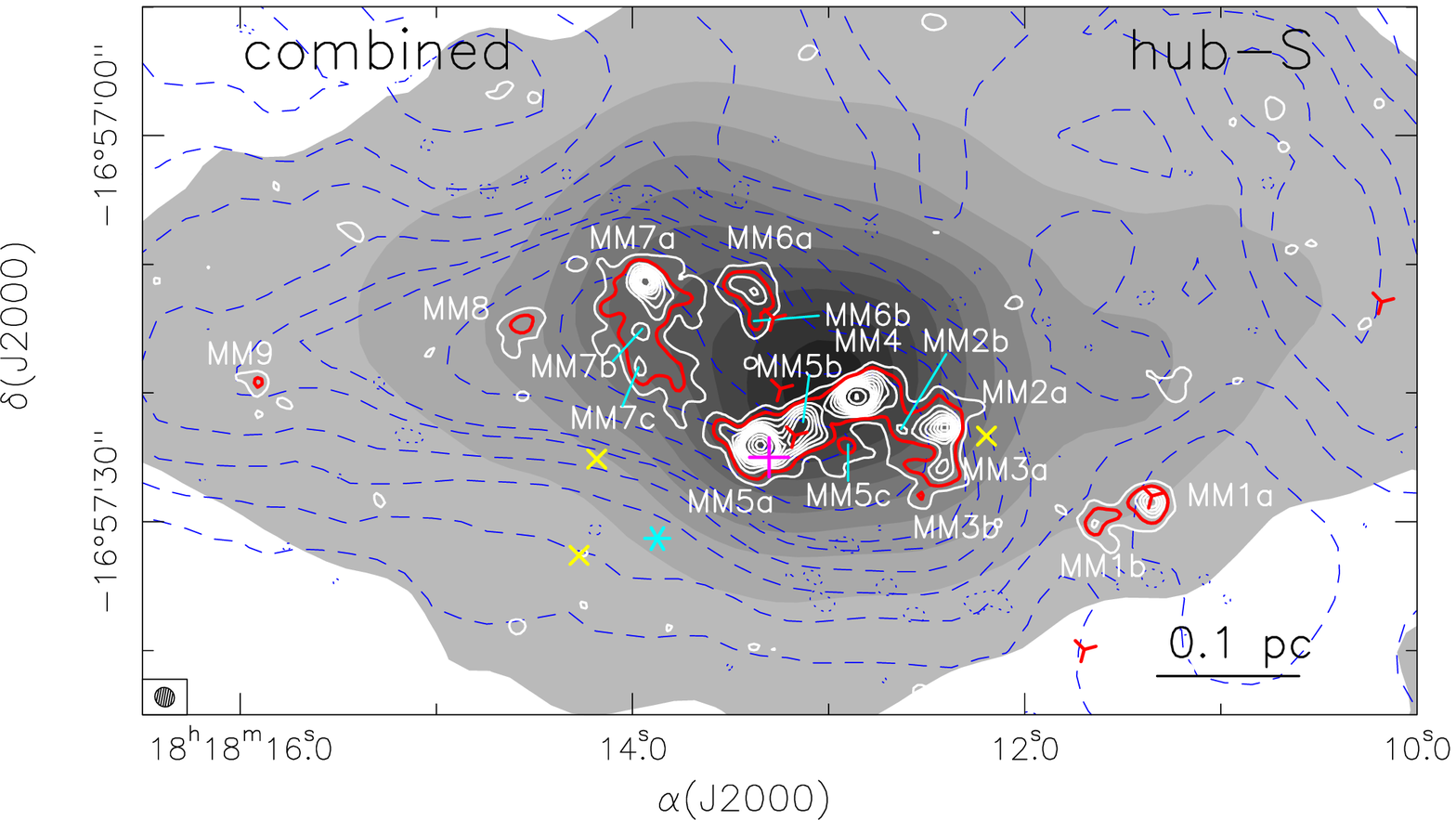,scale=0.7,angle=0} 
	\end{tabular}
\caption{\emph{Top:} Same as Fig.~\ref{fsmacont-hubN} but for G14.2-hub-S. Contours at 1.3~mm start at 3$\sigma$ and increase in steps of 3$\sigma$, where $\sigma$ is the rms of the map listed in Table~\ref{tsmaobs}. 
\emph{Bottom:} Contour map of the SMA dust emission at 1.3~mm toward G14.2-hub-S obtained by combining the compact and extended configuration observations (angular resolution $\simeq1\farcs5$) overlaid on the dust emission at 350~\mum\ from the CSO SHARC~II bolometer (grey scale) and the \nh\,(1,1) integrated intensity map (dashed contours)  from \citet{busquet2013}. 
Contours at 1.3~mm range from 3 to $30\sigma$ increasing in steps of  $3\sigma$, and from 30 to $60\sigma$ in steps of  $10\sigma$, where $\sigma$ is the rms of the map listed in Table~\ref{tsmaobs}. Blue dotted contour corresponds to the $-3\sigma$ contour level. 
The red thick contour level denotes the $6\sigma$ level used to identify the continuum sources. 
The SMA synthesized beam is shown in the bottom left corner of each image.
Symbols are the same as in Fig.~\ref{g14-nh311}.
}
\label{fsmahubS-comb}
\end{center}
\end{figure*}

The higher angular resolution images ($\sim1\farcs5$) 
reveal that both hubs fragment further.
The resulting 1.3~mm continuum maps are presented in Fig.~\ref{fsmacont-hubN} (right panel) and Fig.~\ref{fsmahubS-comb} (bottom panel), which were obtained using all available SMA configurations.
The sensitivity of the final SMA images is the same: 1~m\jpb. At a $6\sigma$ level, this sensitivity corresponds to a mass sensitivity of 0.7~\mo, assuming a dust temperature of 17~K (see below), a gas-to-dust mass ratio of 100, and a dust mass opacity coefficient at 1.3~mm per unit mass density of dust and unit length of 0.899~\cmd\,g$^{-1}$,  which 
corresponds to coagulated grains with thin ice mantles in cores of densities $\sim10^6$~\cmt\ \citep{ossenkopf1994}.
We then identified sources having peak fluxes above the $6\sigma$ rms noise level. The number of millimeter continuum condensations in hub-N is 9 while in hub-S the number of millimeter condensations is 17. In Table~\ref{tmmsources} we report the identified condensations for each hub, 
listing their position, peak flux, total flux density, and deconvolved size obtained for a 2-dimensional Gaussian fitting, when possible. For those sources that cannot be fitted with a Gaussian function we estimated the flux density by integrating the emission above the 4$\sigma$ level.
We additionally report the association of a dust millimeter source with a near-IR source, and the presence of infrared excess 
at 4.5~\mum\ according to the catalog of \citet{povich2010}, as well as its association with \hho\ maser emission \citep{wang2006}.

As can be seen in the close-up image of hub-N (Fig.~\ref{fsmacont-hubN}-right panel) the high angular resolution map shows that in addition to MM2 and MM3, the strongest millimeter continuum source MM1, which has a peak intensity of  56~m\jpb, 
breaks up in 6 additional sources. The strongest source MM1a is associated with a \hho\ maser spot \citep{wang2006}. 
There are two sources located north of MM1a, labelled as MM1e and MM1f, and three faint sources south of MM1a, labelled as MM1b, MM1c, and MM1d. 
MM1f is associated with a YSO, identified in the \textit{Spitzer} image by \citet{povich2010}, in Stage~0/I of evolution, which means it is dominated by an infalling envelope.  Moreover, this source presents a 4.5~\mum\ excess \citep{povich2010}. Sources with this feature are typically called ``Extended Green Objects'' (EGOs) and is due to protostellar outflow activity, indicating that MM1f could be a massive YSO candidate \citep{cyganowski2008}. We detected an additional source, called MM1g, which lies about $5\farcs6$ south east of MM1a whose emission is very compact and faint.
While MM3 is still detected in the high-angular resolution map, the emission of MM2, detected only at a $4\sigma$ level, is filtered out due to a combination of a faint and flattened structure.

Regarding the high angular resolution 1.3~mm image of hub-S, Fig.~\ref{fsmahubS-comb} (bottom panel) shows that the snake-shape structure located at the center of the image, which consisted in 4 fragments in the compact configuration, breaks up into 7 condensations. The strongest and compact source MM5a is associated with a \hho\ maser \citep{wang2006} and MM5b has its peak position very close to a YSO in the Stage 0/I of evolution that presents an excess in the 4.5~\mum\  \textit{Spitzer} band. MM1a and MM1b lie at the border of the dust envelope, about $15''$ to the west of MM5a. The brightest source, MM1a, is associated with an infrared source classified by \citet{povich2010} as a YSO in the Stage 0/I of evolution (\ie\ dominated by an infalling envelope). MM6 splits in two sources separated $2\farcs2$, and MM7 breaks up in three sources, MM7a is the strongest one, and MM7b and MM7c are $3''$ and $6''$ to the south, respectively. The higher sensitivity of the combined map allowed to detect two additional sources, MM8 and MM9, with a high-enough signal-to-noise ratio. 

Assuming that the dust emission at 1.3~mm is optically thin and that the temperature distribution is uniform, we estimated the mass of each 
millimeter condensation adopting a gas-to-dust ratio of 100 and using the dust mass opacity coefficient at 1.3~mm per unit mass density of dust and unit length of 0.899~\cmg\ \citep{ossenkopf1994}. In both hubs, the rotational temperature obtained from \nh\ observations is 15~K \citep{busquet2013}. We converted \Trot\ to gas kinetic temperature using the expression provided  by \citet{tafalla2004}, and obtained \Tkin$\simeq$17~K. Assuming \Tkin=\Td\ the masses of the continuum sources are estimated to be in the range $1-13$~\mo\ in hub-N and $0.7-18$~\mo\ in hub-S. We report the mass of each millimeter condensation in Table~\ref{tmmsources}.
Due to the uncertainty in the dust mass opacity coefficient,
the values of the derived masses are good to within a factor of 2.
While in hub-N only MM1a has a mass larger than 10~\mo, in hub-S there are four sources with $M>10$~\mo. 
The sum of all mass estimated from the 1.3~mm dust continuum emission is 34.2~\mo\ and 95.1~\mo\ in hub-N and hub-S, respectively. 
It is worth noting that the mass of all millimeter condensations, except for MM2 in hub-N, have been measured using the combined image.

\begin{table*}[!ht]
\caption{SMA 1.3~mm continuum sources for the two hubs}
\centering
\begin{tabular}{l c c c c c c c c c c c}
\hline\hline\noalign{\smallskip}	
  Source\_ID 	&\multicolumn{2}{c}{Position\supa}  &I$_{\nu}^{\mathrm{peak}}$ 		&S$_{\nu}$ 	&\multicolumn{2}{c}{Deconvolved Size}  &P.A. 	&$M_{\rm {H_{2}}}$ &near-IR\supb  	 	 &\hho\,\supc    &4.5~\mum\,\supb    \\
\cline{2-3}
\cline{6-7}
		&$\alpha$(J2000.0) 	&$\delta$(J2000.0)  &(m\jpb) 				&(mJy)		&(arcsec)			&(AU$\times$AU) 	  &(\degr)	&(\mo) &source? 	 &maser? &excess?\\
\hline
\textbf{Hub-N} \\
\hline
MM1a 	&18:18:12.32 		&$-$16:49:28.1  	&56.4$\pm$0.9			&118.3$\pm$2.7    &0.8$\times$0.7	&1630$\times$1380	&\phn99	&13.0	&\ldots  	&$\surd$&\ldots \\ 
MM1b\supe\ &18:18:12.30        &$-$16:49:28.7         &\phn9.2$\pm$0.2		&\phnn9.6$\pm$0.8  &\ldots	&$<2890$	&\ldots		&\phn1.1	&\ldots	&\ldots&\ldots \\
MM1c\supe\	&18:18:12.32 		&$-$16:49:31.9  	&\phn6.6$\pm$0.4			&\phn10.0$\pm$2.5    &\ldots	&$<2890$		&\ldots		&\phn1.1	&\ldots	&\ldots&\ldots \\
MM1d\supe\	&18:18:12.33   &$-$16:49:30.5		&\phn6.7$\pm$0.5			&\phn10.1$\pm$1.8	 &\ldots	&$<2890$	&\ldots		&\phn1.1	&\ldots	&\ldots&\ldots	 \\
MM1e\supe\ 	&18:18:12.37 		&$-$16:49:26.9  	&11.0$\pm$1.2			&\phn25.2$\pm$0.7    &\ldots	&$<2890$		&\ldots		&\phn2.8	&\ldots	&\ldots&\ldots\\
MM1f 	&18:18:12.33		&$-$16:49:25.3		&14.0$\pm$0.8		&\phn59.2$\pm4.5$	&2.5$\times$0.8	&4950$\times$1650	&111		&\phn6.5	&$\surd$\ldots&$\surd$ \\
MM1g		&18:18:12.71 		&$-$16:49:30.1		&\phn6.4$\pm$0.9		&\phnn9.2$\pm$0.3		&0.8$\times$0.3	&1630$\times$\phn510	&\phn83	&\phn1.0	&\ldots	&\ldots&\ldots\\
MM2\supd\ &18:18:12.67 &$-$16:50:14.3	 	&16.9$\pm$2.2			&\phn33.1$\pm$6.3 &4.1$\times$2.6	&8120$\times$5150	&\phn68		&\phn3.6		&\ldots		&\ldots     &\ldots \\
MM3 &18:18:13.05 		&$-$16:49:41.3         &18.1$\pm$0.9			&\phn36.2$\pm$2.6 &0.9$\times$0.6	&1780$\times$1190	&\phn97		&\phn4.0	&\ldots		&\ldots	&\ldots \\
\hline
\textbf{Hub-S} \\
\hline
MM1a &18:18:11.36 &$-$16:57:28.6	&17.4$\pm$0.2			&\phn44.3$\pm$0.8 &2.0$\times$1.7	&3920$\times$3360	&119		&\phn4.9	&$\surd$	&\ldots	&\ldots \\ 
MM1b &18:18:11.61 &$-$16:57:29.8	&\phn8.6$\pm$0.2		&\phn28.9$\pm$0.9 &2.9$\times$1.6	&5760$\times$3170	&103		&\phn3.2	&\ldots		&\ldots	&\ldots \\
MM2 &18:18:12.43 &$-$16:57:22.6	&23.1$\pm$0.3			&\phn83.3$\pm$1.0 &2.5$\times$2.2	&4960$\times$4360	&117		&\phn9.2	&\ldots		&\ldots	&\ldots\\
MM2b\supe &18:18:12.63 &$-$16:57:22.8 &\phn9.6$\pm$0.3	&\phnn8.8$\pm$1.7   &\ldots		&$<2800$			&\ldots	&\phn1.0	&\ldots		&\ldots	&\ldots  \\
MM3a &18:18:12.42 &$-$16:57:25.7	&12.4$\pm$0.2			&\phn48.9$\pm$1.1 &2.3$\times$2.0	&4550$\times$3960	&168		&\phn5.4	&\ldots		&\ldots	&\ldots \\ 
MM3b\supe &18:18:12.52 &$-$16:57:27.9 &\phn6.3$\pm$0.2	&\phnn6.2$\pm$1.5	&\ldots		&$<2800$			&\ldots	&\phn0.7	&\ldots		&\ldots	&\ldots  \\
MM4 &18:18:12.84 &$-$16:57:20.3	&43.6$\pm$0.2			&136.6$\pm$0.9 	  &2.8$\times$1.5	&5600$\times$2970	&111	&15.1	&\ldots		&\ldots	&\ldots\\
MM5a &18:18:13.34 &$-$16:57:24.1	&56.1$\pm$0.2			&161.2$\pm$0.8 	  &2.2$\times$1.8	&4430$\times$3530	&\phn93	&17.8	&\ldots		&$\surd$&\ldots\\
MM5b &18:18:13.15 &$-$16:57:22.8	&26.2$\pm$0.3			&105.2$\pm$1.1 	  &3.4$\times$1.8	&6670$\times$3600	&148		&11.6	&$\surd$	&\ldots &$\surd$\\ 
MM5c\supe &18:18:12.90 &$-$16:57:24.0  &\phn7.9$\pm$0.4	&\phnn7.6$\pm$0.5	&\ldots		&$<2800$				&\ldots	&\phn0.8	&\ldots		&\ldots	&\ldots	 \\
MM6a &18:18:13.40 &$-$16:57:12.2	&11.3$\pm$0.3			&\phn59.9$\pm$1.4 &3.5$\times$2.6	&6940$\times$5240	&\phn29	&\phn6.6	&\ldots		&\ldots	&\ldots\\
MM6b\supe &18:18:13.38 &$-$16:57:14.4  &\phn7.3$\pm$0.5	&\phnn6.6$\pm$0.8	&\ldots	&$<2800$				&\ldots	&\phn0.7	&\ldots		&\ldots	&\ldots \\
MM7a &18:18:13.92 &$-$16:57:11.7	&25.6$\pm$0.2			&\phn94.3$\pm$1.1    &2.9$\times$1.9	&5750$\times$3880	&\phn28 &10.4	&\ldots		&\ldots	&\ldots \\
MM7b\supe &18:18:13.95 &$-$16:57:15.0	&10.1$\pm$0.7		&\phn19.5$\pm$1.2 &\ldots		&$<2800$			&\ldots	&\phn2.1	&\ldots		&\ldots	&\ldots\\ 
MM7c\supe &18:18:13.97 &$-$16:57:18.0	&\phn9.9$\pm$1.8	&\phn14.9$\pm$1.5 &\ldots		&$<2800$			&\ldots	&\phn1.6	&\ldots		&\ldots	&\ldots\\
MM8 &18:18:14.86 &$-$16:57:14.9	&\phn7.3$\pm$0.2		&\phn25.8$\pm$1.0 &2.9$\times$1.8	&5870$\times$3490	&131		&\phn2.8	&\ldots		&\ldots	&\ldots	\\
MM9 &18:18:15.93 &$-$16:57:19.2	&\phn5.8$\pm$0.3		&\phn11.3$\pm$0.6 &2.0$\times$0.9	&3930$\times$1750	&\phn60	&\phn1.2	&\ldots		&\ldots	&\ldots\\
\hline
\end{tabular}
\begin{list}{}{}
\item[\supa]{Source positions determined from a 2D Gaussian fit. Units of R.~A in (h:m:s) and Decl. in ($\degr$:$'$:$''$).}
\item[\supb]{near-IR source or 4.5~\mum\ \textit{Spitzer} excess reported by \citet{povich2010}.}
\item[\supc]{\hho\ maser detected \citep{wang2006}.}
\item[\supd]{Properties of MM2 determined using the SMA image at $4''$ resolution.}
\item[\supe]{A 2D gaussian cannot be fitted and the peak intensity and total flux density were obtained by integrating the emission within the 
4$\sigma$ contour level.}
\end{list}
\label{tmmsources}
\end{table*}

\subsection{Fragmentation level}

\begin{figure*}[!ht]
\begin{center}
\begin{tabular}[t]{cc} 
\epsfig{file=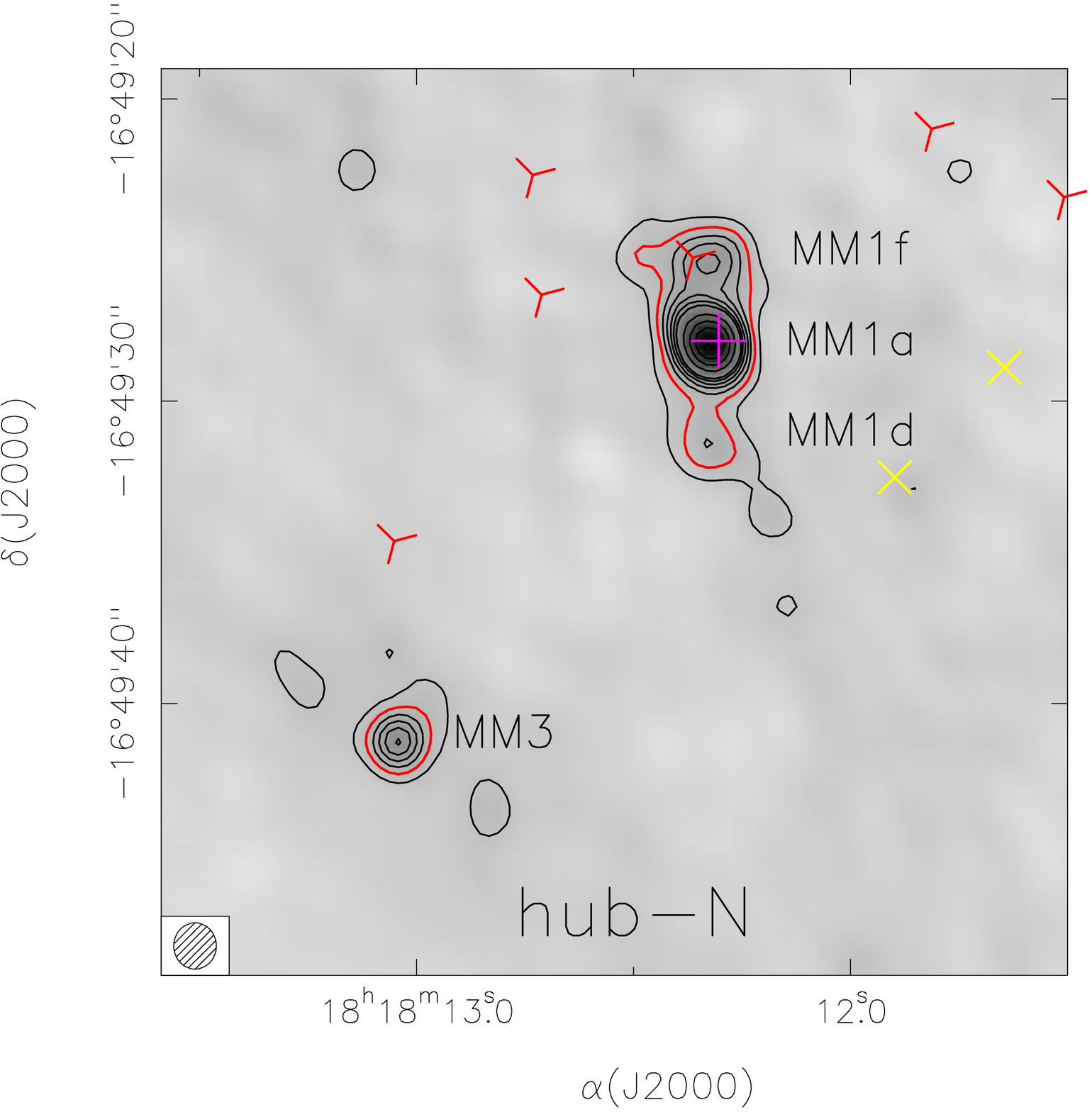,scale=0.43,angle=0} &
\epsfig{file=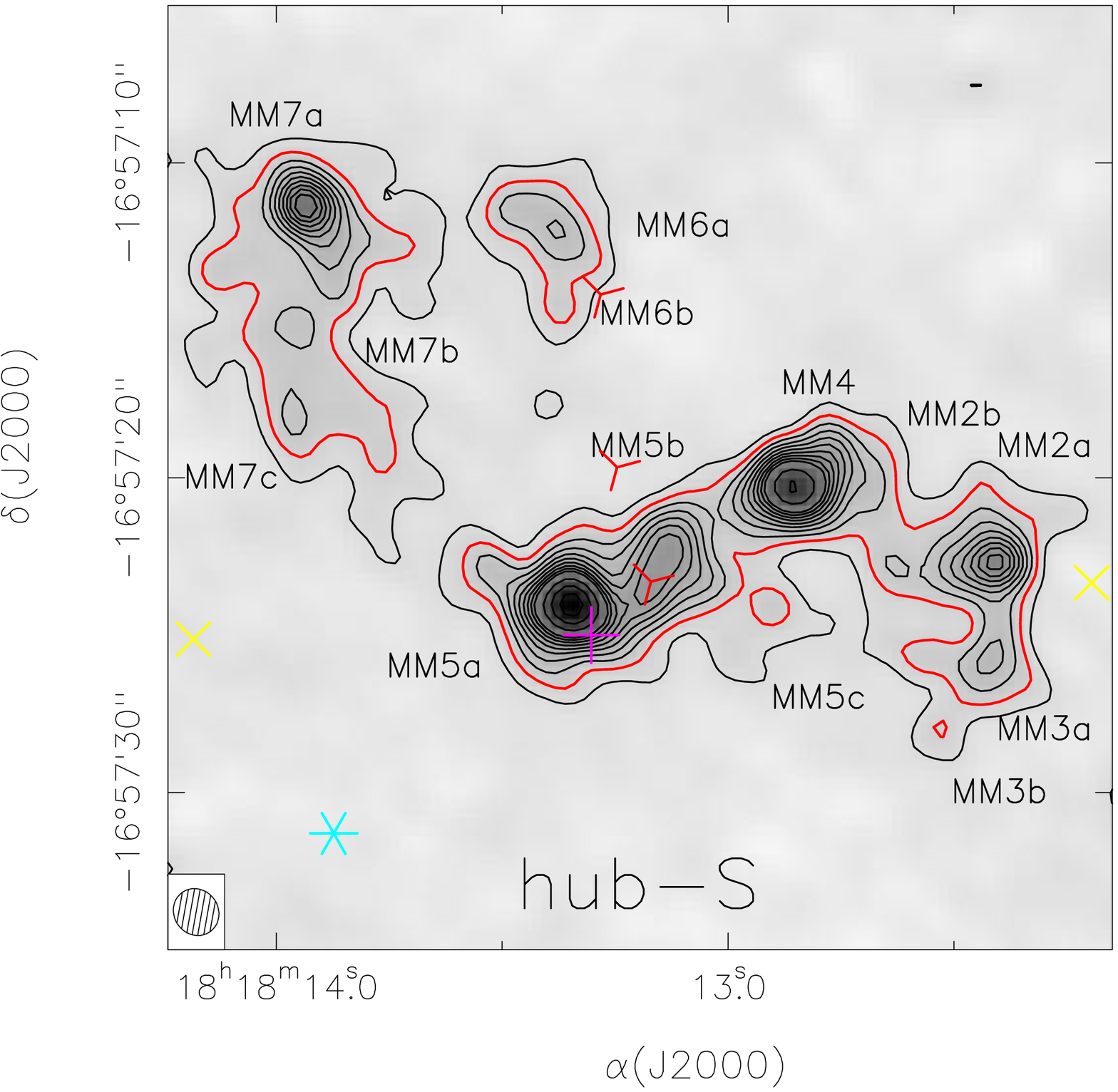,scale=0.43,angle=0} 
	\end{tabular}
\caption{Contour map of the SMA dust emission at 1.3~mm toward hub-N (left) and hub-S (right) showing a field of view of $\sim30"\sim0.3$~pc. Note that for hubN the combined map has been convolved to the same beam as the hub-S image for an easy comparison of the fragmentation level in each hub. The synthesized beam, $1\farcs53\times1\farcs41$, is shown in the bottom left corner of each image. 
Contours and symbols are the same as in Fig.~\ref{fsmacont-hubN} and Fig.~\ref{fsmahubS-comb}. The 6$\sigma$ rms level is indicated by the red thick contour.
}
\label{fmmreg}
\end{center}
\end{figure*}

In the previous sections we have shown the SMA maps at different resolutions. Now, the aim is to estimate the fragmentation level in each hub. To do so, we need first to obtain images with the same angular resolution to be easily comparable. We then performed the imaging of each hub using a common $uv$-range of visibilities (8-160~k$\lambda$). The observations of hub-N have more visibilities at longer $uv$-distances than hub-S, which results in a slightly smaller beam, especially in one direction (roughly the north-south direction), so we convolved the map to the beam obtained for hub-S (\ie\ $1\farcs53\times1\farcs41$). 
In addition, since our SMA observations consist of a small mosaic instead of a single pointing, and the total field of view of each hub is different, we need to define a region in which we can evaluate in a consistent way the fragmentation level in each hub. The typical radii of compact and embedded clusters and subclusters lie in the range $0.1-0.2$~pc \citep[\eg][]{testi1999,alexander2012,kuhn2014}. 
Then, we defined a radius of 0.15~pc, which corresponds to a region of $\sim30''$ at the distance of the source. 
In Fig.~\ref{fmmreg} we present the 1.3~mm continuum emission in a field of view of 30$''$ for hub-N (left panel) and hub-S (right panel). Adopting this field of view and the $6\sigma$ detection threshold, we found 4 fragments in hub-N and 13 fragments in hub-S. These values are listed in Table~\ref{tquant}. Therefore, using the proper comparison we still clearly see that hub-S has a higher fragmentation level than hub-N.

\section{Analysis \label{sanal}}

\subsection{Radial Intensity Profiles \label{model}}

Recently, \citet{palau2014} studied the relation between the
fragmentation level and the density structure in a sample of 19
massive dense cores selected to be in a similar evolutionary stage, \ie\ all regions harbor intermediate/high-mass protostars deeply embedded in massive dense cores that have not yet developed an ultra-compact \hii\ region, hence comparable to our targets hub-N and hub-S. 
They find a weak (inverse) trend of fragmentation level and density power-law index, with steeper density profiles tending to show lower 
fragmentation and vice-versa. One of the main results of this work is that, within a given radius, the fragmentation level increases with average density as a combination of flat density profile and high central density. 
A comparison with magnetohydrodynamic simulations
\citep{commercon2011} suggests that the cores showing no fragmentation
could be related with a strong magnetic field. 
In a follow-up work using the same sample of massive dense cores, \citet{palau2015} analyze whether the fragmentation level can be explained by turbulent or thermal support, and conclude that the observed fragmentation level is more consistent with pure thermal Jeans fragmentation than fragmentation including turbulent support.

\begin{figure*}[!ht]
\begin{center}
\begin{tabular}[t]{c} 
\epsfig{file=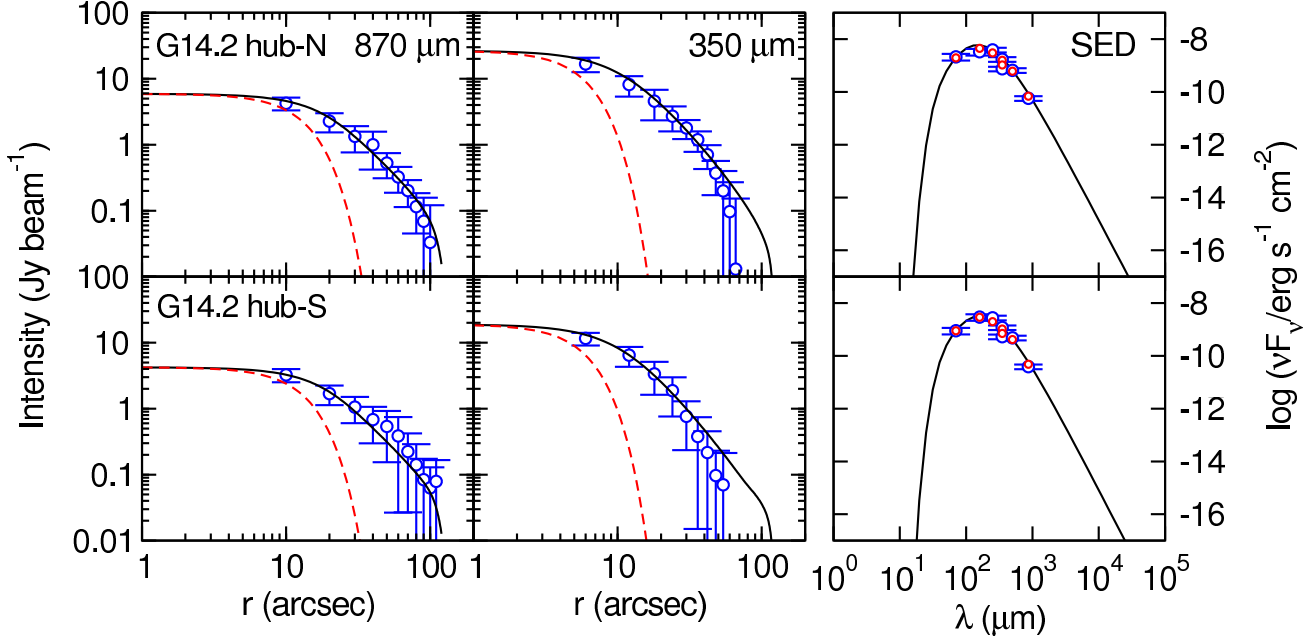,scale=0.9,angle=0} 
	\end{tabular}
\caption{Best fits for hub-N (top panels) and hub-S (bottom panels). The left
and middle panels show the radial intensity profiles at 870 and 350~\mum. Open
blue circles with error bars show the data, the solid black line shows the
best-fit  model, and the dashed red line shows the beam profile. The right panel
shows the spectral energy distribution. Blue open circles with error bars show
the observed fluxes, and the open red circles show the model values integrated
for the same  aperture than the observation. The continuous black line is the
SED integrated up to the model envelope radius $r_\mathrm{max}$ (see Table
\ref{tfitpar}). Note that SHARC~II and LABOCA cannot detect structures $>1'$ scale. 
}
\label{fradprofile}
\end{center}
\end{figure*}

In order to investigate whether the differences of the fragmentation level in
hub-N and hub-S could be explained in terms of the density structure of the envelope, we applied
the model presented in \citet{palau2014} to fit simultaneously the radial
intensity profile at 870~\mum\ and 350~\mum\ and the spectral energy distribution
(SED) assuming spherical symmetry.  
It is important to mention that both hubs are embedded in
filamentary structures, and thus the radial profile along
the filament's main axis has contributions from the filament and the hub. Since
we are interested in modelling the structure of the hub, not the filament, we
extracted the radial profiles in the direction perpendicular to the filament
main axis to avoid contamination from the filament itself at large distances.
The radial intensity profiles at 870~\mum\ and 350~\mum\ were obtained in rings of
10$''$ and 6$''$ width, respectively, as a function of the projected distance from
the clump center.
The radial intensity profiles of hub-N and hub-S at 870~\mum\ and 350~\mum\ are presented
in Fig.~\ref{fradprofile}. 

The SED of each hub was built considering the flux density at 350 and 870~\mum\
from this work, and the flux densities measured by the ESA \textit{Herschel Space
Observatory} \citep{pilbratt2010} in the framework of the Herschel infrared
Galactic Plane Survey (Hi-GAL) project, which performed an unbiased photometric
survey of the Galactic plane in five photometric bands: 70, 160, 250, 350, and
500~\mum\ \citep{molinari2010}. The \textit{Herschel} fluxes have been measured
following the standard procedures described in the Hi-GAL articles
\citep{molinari2010a,elia2013} using CuTEx \citep{molinari2011} by means of a 2-D
Gaussian fit of the source brightness profile
independently at each band.  The fluxes at 870~\mum\ and 350~\mum\ obtained with
the APEX and CSO telescopes, respectively, were estimated within a radius of
22$''$ and 10$''$, respectively. At 350~\mum\ we used both the \textit{Herschel}
and the CSO values. 

In the present model, we describe the density structure using a Plummer-like
function of the form 
\begin{equation}
\rho(r)=\rho_{c}\left[1+\left(\frac{r}{r_{c}}\right)^2\right]^{-p/2}
\end{equation}
where $\rho_c$ is the central density, $r_c$  is the radius of the flat inner
region, and $p$ is the asymptotic power index. 

The temperature was assumed to be a power-law of radius 
\begin{equation}
T=T_{0}\Big(\frac{r}{r_{0}}\Big)^{-q}, 
\end{equation}
where $q=2/(4+\beta)$. We assumed a dust opacity law as a power-law of frequency
with index $\beta$, $\kappa_{\nu}=\kappa_{0}(\nu/\nu_{0})^{\beta}$, where
$\nu_{0}$ is an arbitrary reference frequency. At $\nu_{0}=230$~GHz, we used the
dust mass opacity coefficient adopted in Sect.~\ref{smahigh}.

We adopted a constant temperature of 8~K for radii larger than the radius where
the temperature distribution of index $-q$ drops to 8~K. We stress that a minimum temperature of 10~K does not alter the results.
Concerning the density
distribution, we adopted, as a maximum radius of the envelope, the radius for which
the envelope density achieved a value comparable to that of the ambient gas of the intecore medium, taken to be
$5\times10^3$~\cmt, the same value adopted in \citet{palau2014}.

The model has five free parameters: 
the dust emissivity index $\beta$, 
the envelope temperature $T_0$ at the reference radius $r_0$, 
the envelope density $\rho_0$ extrapolated to the reference radius $r_0$,  
the radius of the inner region where density is flat (in units of $r_0$) 
$r_c/{r_0}$, and 
the density power law index $p$. 
The reference radius $r_0$ is arbitrary and has been taken as $r_0=1000$~AU. 
It is important to remark that $\rho_0$ is not the actual density at $r_0$, but
the value given by the asymptotic density power law, 
$\rho_0=\rho_c(r_0/r_c)^{-p}$. 
In fact, if $r_0<r_c$, the density $\rho_0$ is higher than the maximum
density $\rho_c$. 
Only in the case  $r_0\gg r_c$ does the density $\rho_0$ correspond to the
actual density at $r=r_0$. 

We used this model to fit simultaneously the observed radial intensity profiles at
870~\mum\ and 350~\mum\, and the observed SED. We adopted an uncertainty in the
radial intensity profiles at both 870 and 350~\mum\ of 20~\%. The uncertainty of
the flux densities used in the SED was 20~\% at all wavelengths, plus an
additional 20~\%, added quadratically, for the 70 and 160~\mum\ values (the
PACS data). Note  that SHARC~II and LABOCA instruments cannot detect structures larger than $1'$ scale.

For each set of model parameters we computed the intensity  map at 870~\mum\ and
350~\mum\, and convolved it with a  Gaussian of the size of the beam of the telescope
used in the observation. The intensity profile was obtained
from the convolved map, and compared with the circularly averaged observed radial intensity profiles.
On the other hand, in order to compare the model with the observed SED, we
computed the model flux density by integration of the model intensity profile
within the  aperture used to estimate the flux for each data point. 

The initial search ranges for the 5 parameters were 
$\beta=1.5\pm1.5$,
$T_{0}=300\pm300$~K, 
$\rho_{0}=(4.0\pm4.0)\times10^{-16}$~g\,\cmt,
$r_{c}/r_{0}=20\pm20$, and 
$p=1.5\pm1$. 
The search was carried out for 13 loops, each loop consisting in 4000 samples of
the parameter space, with a search range reduced a factor 0.75 around the best-fit
value of the parameters found for the last loop. 
A second search was performed with the best values found in the first run as
starting values of the fit parameters. For this second run the initial search
ranges were
$0.2$, $20$~K, $2.0\times10^{-16}$~g\,\cmt, $8$, and $0.15$, 
for $\beta$, $T_0$, $\rho_{0}$, $r_{c}/r_{0}$, and $p$, 
respectively, and the search range was reduced a factor 0.8 for each loop.  We
refer to the work of \citet{palau2014} for further details on the model and the
fitting procedure used to find the best-fit model \citep[see also][for details on
the $\chi^2$ minimization]{estalella2012,sanchez-monge2013}.

\begin{table*}[!t]
\caption{Best-fit parameters to the radial intensity profiles and SED}
\centering
\begin{tabular}{l c c c c c c c c c c c}
\hline\hline\noalign{\smallskip}	
&
&$T_0$\supa 
&$\rho_0$\supa	
&
&	
&
&$\rho_c$\supb	
&
&$r_\mathrm{8K}$\supb 
&$r_\mathrm{max}$\supb \\
Source 
&$\beta$\supa 					
&(K)				
&(g\,\cmt)			
&$r_c/r_0$\supa						
&$p$\supa		
&$\chi_{r}$\supa	
&(g\,\cmt)			
&$q$\supb 	
&(pc)	
&(pc) \\
\hline
hub-N &$1.81\pm0.08$ &$51\pm2$ &$(1.3\pm0.2)\times10^{-15}$ &$21\pm4$ &$2.24\pm0.04$ &0.69 &$(1.4\pm0.6)\times10^{-18}$ &0.34 &1.05 &0.62 \\
hub-S &$1.89\pm0.08$ &$45\pm2$ &$(1.0\pm0.2)\times10^{-15}$ &$20\pm3$ &$2.24\pm0.04$ &0.76 &$(1.2\pm0.5)\times10^{-18}$ &0.34 &0.79 &0.57 \\
\hline
\end{tabular}
\begin{list}{}{}
\item[\supa]{Free parameter fitted by the model: 
$\beta$ is the dust emissivity index; 
$T_0$ and $\rho_0$ are the temperature and density at the reference radius
$r_0=1000$~AU; 
$r_c/r_0$ is the radius of the flat inner region in units of $r_0$;
$p$ is the asymptotic power index;
$\chi_r$ is the reduced chi-square, $[\chi^2_\mathrm{min}/(n-5)]^{1/2}$.}
\item[\supb]{Parameters derived from the modeling:
$\rho_c$ is the central density, $\rho_c=\rho_0(r_c/r_0)^{-p}$;
$q$ is the temperature power-law index; 
$r_{\rm{8K}}$ is the radius of the clump where the temperature
drops to 8~K; 
$r_{\rm{max}}$ is the radius at the assumed ambient density of
$5\times10^3$~g\,\cmt.}
\end{list}
\label{tfitpar}
\end{table*}

Figure~\ref{fradprofile} shows the best fit to the 870 and 350~\mum\ radial
intensity profile and the SED for hub-N (top panel) and hub-S (bottom panel).
The fitted parameters are reported in Table~\ref{tfitpar}.  Interestingly,
despite the fragmentation level being significantly higher in hub-S than in hub-N,
Table~\ref{tfitpar} shows that the parameters inferred from the model are
remarkably similar. The dust emissivity index is $\sim1.8$--1.9, which
corresponds to a temperature power law index of $q\simeq0.34$, consistent with internal heating. The fitted
temperature at 1000~AU is 51$\pm2$~K and 45$\pm2$~K in hub-N and hub-S, respectively.  The
central density, $\rho_c$, is (1.0--1.3)$\times10^{-18}$~g\,\cmt\ and the radius of
the flat inner region is $r_c\simeq0.1$~pc. In Table~\ref{tfitpar} we also show
the radius where the density reaches $5\times10^3$ ~\cmt\ and the radius where the
temperature drops down to 8~K.  In both hubs we obtained the same index
$p\simeq2.2$ of the asymptotic density power law. 


\subsection{General properties of the hubs \label{aprop}}

Once we obtained the best-fit model, we derived different physical quantities, which are reported in Table~\ref{tquant}. We estimated the total mass of each hub by integrating the model intensity up to the radius where the density profile could be measured. 
Within the errors, the mass of both hubs is essentially the same, being slightly more massive hub-N than hub-S. Given that we evaluated the fragmentation level within a region of 0.15~pc in radius, we estimated the mass ($M_{\rm{0.15\,pc}}$), average density ($n_{\rm{0.15\,pc}}$), and average temperature ($T_{\rm{0.15\,pc}}$) within a radius of 0.15~pc. As expected, the inferred values in both hubs are remarkably similar. Note that $T_{\rm{0.15\,pc}}$ is 18~K and 16~K in hub-N and hub-S, respectively, which is consistent with the kinetic temperature, \Tkin=17~K, derived using the \nh\ data \citep{busquet2013}.

\begin{table*}[ht]
\begin{scriptsize}
\caption{Properties of the two Hubs}
\centering
\begin{tabular}{l c c c c c c c c c c c c c c}
\hline\hline\noalign{\smallskip}	
		&		&		&$M_{\rm{tot}}$\supc	&\lbol\supc	&$M_{\rm{0.15\,pc}}$\supc	&$n_{\rm{0.15\,pc}}$\supc		&$T_{\rm{0.15\,pc}}$\supc 	&$M_{\rm{Jeans}}^{\rm{th}}$\supd	&$\sigma_{\rm{1D,nth}}$\supe	&$M_{\rm{Jeans}}^{\rm{tot}}$\supd	
&	&	&		& \\	
Source 	&$N_{\rm{mm}}$\supa		&$N_{\rm{IR}}/N_{\rm{mm}}$\supb	&(\mo)	&(\lo)	&(\mo)	&($\times10^5$~\cmt)				&(K)					&(\mo)				&(\kms)				&(\mo)		&$\mathcal{M}$\supf		&$\mathcal{M}_{\rm{A}}$\supf 		&$\lambda$\supf 		&$\beta_{\rm{rot}}$\supf 	 \\
\hline
hub-N 		&\phn4				&1.7		&$979\pm329$		&995 	&126		&1.3	&18	&1.33	&0.83		&70		
&5.6		&0.4				&0.9		&0.016 \\
hub-S 		&13					&0.4		&$717\pm250$		&531 	&105		&1.1	&16	&1.21	&0.88		&90		
&6.4 		&0.3				&0.5		&0.015 \\
\hline
\end{tabular}
\begin{list}{}{}
\item[\supa]{$N_{\rm{mm}}$ is the number of fragments obtained within a field of view of 0.3~pc of diameter (see Section 3.4).}
\item[\supb]{Ratio between the number of IR sources without a millimeter counterpart within $\sim0.3$~pc, obtained from the catalog of \citet{povich2010}, over the number of millimeter sources detected in this work.}
\item[\supc]{Parameters inferred from the modeling. $M_{\rm{tot}}$ is the mass computed from the model by integrating up to the radius where the density profile could be measured; \lbol\ is the bolometric luminosity obtained by integration of the SED;
$M_{\rm{0.15\,pc}}$, $n_{\rm{0.15\,pc}}$, and $T_{\rm{0.15\,pc}}$ are the mass, average density, and average temperature within a radius of 0.15~pc. }
\item[\supd]{$M_{\rm{Jeans}}^{\rm{th}}$ is the thermal Jeans mass for the temperature $T_{\rm{0.15\,pc}}$ and density $n_{\rm{0.15\,pc}}$ (Eq.~6 of \citealt{palau2015}); $M_{\rm{Jeans}}^{\rm{tot}}$ is the total Jeans mass using the total (thermal + non-thermal) velocity dispersion (Eq.~4 of Palau \et\ 2015). }
\item[\supe]{$\sigma_{\rm{1D,nth}}$ is the non-thermal turbulent component of the velocity dispersion estimated as $\sigma_{\rm{1D,nth}}=\sqrt{\sigma_{\rm{obs}}^2-\sigma_{\rm{th}}^2}$ using the \nh\,(1,1) data.}
\item[\supf]{ $\mathcal{M}$ is the Mach number defined as $\sigma_{\mathrm{3D,nth}}/c_\mathrm{s}$ where $\sigma_{\mathrm{3D,nth}}=\sqrt3\,\sigma_{\mathrm{1D,nth}}$; $\mathcal{M}_{\rm{A}}$ is the Alfvén Mach number, expressed as   $\sqrt3\sigma_{\mathrm{1D,nth}}/v_{A}$ where $v_{A}=B_\mathrm{{tot}}/\sqrt{4\pi\rho}$ is the Alfvén speed; $\lambda=(M/\Phi_{\rm{B}})/(M/\Phi_{\rm{B}})_{\rm{cr}}$ is the mass-to-magnetic-flux ratio; and $\beta_{\rm{rot}}$ is the rotational-to-gravitational energy ratio as defined in \citet{chen2012} and derived using \nh\,(1,1) data.}
\end{list}
\label{tquant}
\end{scriptsize}
\end{table*}

Given that our aim is to investigate the interplay between the different physical agents acting during the fragmentation process, 
in the following we estimated some relevant physical quantities to evaluate the relative importance of the different forces within a region of 0.3~pc in diameter, the region where we assessed the fragmentation level in each hub. \\

\textit{Velocity dispersion.}
We extracted the averaged \nh\,(1,1) spectrum of both hubs and fitted the hyperfine structure using the CLASS package of the GILDAS software to measure the \nh\,(1,1) line width, $\Delta$$v_{\rm{obs}}$, over a region of 0.3~pc of diameter (\ie\ 30$''$). Within this region, infall and rotation motions may contribute to the observed linewidth, which is 2.27 and 2.38~\kms\ in hub-N and hub-S, respectively. Thus, in order to obtain the contribution due to turbulent motions we estimated the velocity gradient within this region of 0.3~pc using the \nh\ data and found 3.8~\kms\,pc$^{-1}$, which can be attributed to rotation and/or infall motions (see the derivation of the rotational-to-gravitational energy ratio for further details) and subtract, in quadrature, this contribution from the observed line width. The resulting values are 1.96 and 2.09~\kms\ in hub-N and hub-S, respectively.  We subsequently obtained the observed velocity dispersion, $\sigma_{\rm{1D,obs}}$, and separate the thermal ($\sigma_{\rm{1D,th}}$) from the non-thermal turbulent ($\sigma_{\rm{1D,nth}}$) contribution of the observed velocity dispersion following the procedure described in \citet{palau2015} and using the average kinetic temperature within a radius of 0.15~pc, which has been obtained from the model (see Table~\ref{tquant}). Once we derived the thermal contribution of the velocity dispersion, $\sigma_{\rm{1D,th}}=\sqrt{\kappa_{\rm{B}}T/(\mu\,m_{\rm{H}})}$ with $\mu=17$ for \nh, which is $\sim0.1$~\kms, we obtained the non-thermal turbulent component, which is listed in Table~\ref{tquant}. The isothermal sound speed at the temperature of the two hubs is $c_{\rm{s}}\simeq0.24$~\kms\ and the total velocity dispersion, defined as $\sigma_{\rm{1D,tot}}=\sqrt{c_{\rm{s}}^2+\sigma_{\rm{1D,nth}}^2}$, is $\sim0.9$~\kms.\\


\textit{Mach number.}
We computed the Mach number $\mathcal{M}$ as in \citet{palau2015} and obtained a value of 5.6--6.4, which means that hubs show supersonic non-thermal gas motions.\\

\textit{Jeans mass and Jeans number.}
We investigated whether the fragmentation can be explained in terms of pure thermal Jeans fragmentation and including both thermal and non-thermal support adopting Equations~(6) and (4) of \citet{palau2015}, respectively. 
Using the average temperature and density within a radius of 0.15~pc (listed in Table~\ref{tquant}), we derived a thermal Jeans masses of 1.3~\mo\ and 1.2~\mo\ for hub-N and hub-S, respectively. The Jeans mass increases up to values of 70~\mo\ and 90~\mo\ if we include the non-thermal support. The number of expected fragments, $N_{\rm{Jeans}}$, 
is given by the ratio between the mass of each hub inside a region of 0.15~pc in radius, $M_{\rm{0.15\,pc}}$, and the Jeans mass, $M_{\rm{Jeans}}$, scaled by a core formation efficiency (CFE) to take into account that not all the mass of the hub will be accreted and converted into compact fragments
(equation~(7) of \citealt{palau2015}). Thus, to estimate the number of fragments we need first to compute the CFE of each hub, which can be calculated from the ratio of the sum of the masses of all fragments measured with the SMA, listed in Table~\ref{tmmsources}, and the total mass of the hub obtained from the model and listed in column~(4) of Table~\ref{tquant}. The values of the CFE are 3\,\%\ and 13\,\%\ for hub-N and hub-S, respectively, consistent with the values reported in other works \citep[\eg][]{bontemps2010,palau2013,palau2015,louvet2014}. Using the values of CFE, we obtained that the number of expected fragments considering pure thermal Jeans fragmentation is 3 and 12 fragments in hub-N and hub-S, respectively, while if we add turbulent support the expected number of fragments is smaller than unity. Therefore, the observed number of fragments in each hub is in very good agreement with pure thermal Jeans fragmentation. \\

\textit{Alfvén Mach number.}
To evaluate the importance of turbulence against the magnetic field we calculated the Alfvén Mach number, $\mathcal{M}_{\rm{A}}$, defined as $\mathcal{M}_{\rm{A}}=\sqrt{3}\sigma_{\rm{1D,nth}}/v_{\rm{A}}$, where $v_{\rm{A}}=B_{\rm{tot}}/\sqrt{4\pi\rho}$ is the Alfvén speed and $B_{\rm{tot}}$ is the total magnetic field. Recently, Santos \et\ (in preparation) has measured the angular dispersion factor of the magnetic field and the sky-projected magnetic field strength through near-infrared polarimetric data, which trace the magnetic field in the diffuse medium surrounding all filaments and hubs previously identified in \citet{busquet2013}. For the diffuse gas, the magnetic field is 0.3~mG in hub-N and 0.5~mG in hub-S.  If we now assume that we are in the magnetic field controlled contraction regime, $B\propto\,n_{\rm{H_2}}^{1/2}$, the values of the magnetic field extrapolated to the densities reported in Table~\ref{tquant} are 1.0~mG in hub-N and 1.5~mG in hub-S, which yields to $\mathcal{M}_{\rm{A}}$ of 0.4 and 0.3 in hub-N and hub-S, respectively (see Table~\ref{tquant}). Clearly $\mathcal{M}_{\rm{A}}<1$, indicating sub-Alfv\'enic conditions, even if we take into account the uncertainty in the derived values of the sky-projected magnetic field, of the order of 50~\%. Thus, magnetic energy dominates over turbulence. Similar values are found by \citet{pillai2015} towards two massive IRDCs using submillimeter polarization data, and by \citet{franco2010} toward the Pipe nebula using optical data. \\

\textit{Mass-to-magnetic-flux ratio.}
The relevance of magnetic field with respect to the gravitational force can be assessed by the mass-to-magnetic-flux ratio ($M/\Phi$). This ratio is expressed in terms of a critical value as $\lambda=(M/\Phi)/(M/\Phi)_{\rm{cr}}=7.6\times10^{-21}[N_{\rm{H_{2}}}/\rm{cm^{-2}}][B_{\rm{tot}}/\mu G]^{-1}$ \citep{crutcher2004}. For the column density, $N$(\hh), we used the mass obtained from the model within a region of 0.3~pc of diameter. Assuming spherical symmetry $N_{\rm{H_{2}}}=M/(2m_{\rm{H}}\pi\,R^{2})$,  we obtained $N$(\hh)$=1.1\times10^{23}$, and 0.9$\times10^{23}$~\cmd\ in hub-N and hub-S, respectively. The resulting mass-to-flux ratio are 0.9 and 0.5 in hub-N and hub-S, respectively (see Table~\ref{tquant}). Taking into account the uncertainty associated with the sky-projected magnetic field strength, the mass-to-flux ratio is close to unity, as expected, since there is ongoing star formation activity in these hubs, and hence part of the mass has already been accreted onto the protostars. Although global gravitational collapse is certainly expected in these hubs, magnetic field cannot be ignored. \\

\textit{Rotational-to-gravitational energy ratio.}
Finally, we computed the rotational-to-gravitational energy ratio, $\beta_{\rm{rot}}=E_{\rm{rot}}/E_{\rm{grav}}$, following Equation~(1) of \citet{chen2012}. We used the velocity field map of \nh\,(1,1) shown in \citet{busquet2013} and estimated the velocity gradient within a region of 0.3~pc in diameter, finding similar values in both hubs, around 3.8~\kms\,pc$^{-1}$. Both rotation and infall motions contribute to the observed velocity gradient, and gravitational collapse is certainly expected in these two hubs as they contain deeply embedded protostellar objects. However, with the current data we cannot separate the contribution due to rotation from that due to infall motions. Assuming a mass infall rate of $10^{-5}-10^{-3}$~\mo/yr \citep{kirk2013,peretto2013,peretto2014} we derived an infall velocity, $v_{\rm{infall}}=\dot{M}/\pi\,R^{2}\,\rho$, lying in the range of 0.015--0.15~\kms. The lack of information from molecular line observations prevent us to estimate the velocity gradient due to infall or collapse motions, and hence, we derived $\beta_{\rm{rot}}$ assuming that the observed velocity gradient is due only to rotation. The inferred values are the same in both hubs, $\beta_{\rm{rot}}=0.02$, which is within the range of values observed in low- and high-mass star-forming regions \citep[\eg][]{goodman1993,chen2012,palau2013}. 
This value should be regarded with caution as the velocity gradients used to estimate $\beta_{\rm{rot}}$ was attributed as due to rotation motions only but it may be affected by the presence of several components and/or infall/outflow motions. 
Nevertheless, this value is slightly above the threshold value of 0.01 obtained in simulations of a rotating core that could fragment if the rotational energy is large enough compared to the gravitational energy \citep[\eg][]{boss1999}.

\section{Discussion \label{sdiscuss}}

In the previous sections we showed that the two hubs associated with
the IRDC G14.2 seem to be twin hubs. The model applied to obtain the
underlying density and temperature structure of these hubs yield the
same physical parameters, indicating a similar internal structure of
the envelope where the fragments are embedded in, and all the derived
quantities reported in Table~\ref{tquant} are notably similar as
well. However, they present different level of fragmentation, hub-S
being more fragmented than hub-N. With the aim of exploring whether these two similar hubs undergo similar fragmentation process, in this section 
we will discuss the interplay between density structure, turbulence, magnetic field, UV radiation feedback, and core evolution in the fragmentation process.

\subsection{Density structure, turbulent fragmentation, and magnetic field}

One of the main results of \citet{palau2014}, who studied the relation between the fragmentation level and the density
structure in a sample of 19 massive dense cores, is that
there is a clear trend of fragmentation level increasing with average density within a given radius as a result of flat density profile and high central density (see Fig.~6 of \citealt{palau2014}), indicating that density structure plays an important role in the fragmentation process. In \citet{palau2014} the massive dense cores were observed
with interferometers reaching spatial scales down to 1000~AU,
picking up only the most compact fragments, and the region
used to evaluate the fragmentation level was 0.05~pc in radius. Apparently, our results seem to be in contradiction with \citet{palau2014} since hub-N and hub-S show the same density structure but they clearly present different fragmentation level, and according to \citet{palau2014} we should expect higher central density and a steeper density profile in hub-N as it presents a lower fragmentation level. However, in the present work we assessed the fragmentation level in a region of $30''$ (\ie\ 0.15~pc in radius) and the SMA combined data is sensitive to structures of 3000-10000~AU, and hence sensitive to  flattened condensations. 
Therefore, to properly compare our result with the work of \citet{palau2014} we evaluated the fragmentation level within the same field of view of \citet{palau2014}, \ie\ 0.1~pc of diameter, and we obtained 3--4 fragments in each hub. From the results of the best-fit model obtained in Section~\ref{model} the density at 0.05~pc is 2.9$\times10^5$~\cmt\ and 2.4$\times10^5$~\cmt\ in hub-N and hub-S, respectively. Including these numbers in Fig.~6(c) of \citet{palau2014} we can see that the observed number of fragments in hub-N and hub-S, $N_{\rm{mm}}\simeq3-4$, is in good agreement with the trend observed by \citet{palau2014}, and the fragmentation level under these conditions would be the same. 
This result suggests that the fragmentation of most compact structures (which will probably form stars in the future) depends on the density structure while fragmentation including larger scales structures ($\gtrsim5000$~AU) might be determined by different processes. 
Thus, we conclude that the differences in the fragmentation level at the scales measured in this work cannot be produced by density and temperature structure of the hub, given that the model results are similar.

The Jeans analysis performed in Section~\ref{aprop} revealed that the fragmentation process can be explained without invoking turbulent support. In fact, as can be seen in Table~\ref{tmmsources}, the most massive cores have masses $>10$~\mo\, which are closer to the turbulent Jeans mass rather than thermal Jeans mass. On the other hand, most cores have masses $\sim1$~\mo\ and are close to thermal Jeans mass, which is consistent with the results of \citet{zhang2015} and \citet{palau2015} regarding the formation of massive dense cores and low-mass fragments, respectively. 
Both the Jeans mass of most cores and the expected number of fragments point towards thermal Jeans fragmentation since the inclusion of turbulence as an additional form of support implies the formation of 4-5 fragment at most and a CFE higher than 100~\%\ (see \citealt{palau2015} for further discussion on CFE). 
Despite internal motions in both hubs belong to the supersonic regime, 
the level of turbulence at these scales (from 0.01 to 0.15~pc) does not seem to play an important role in the fragmentation of these hubs. In particular, the non-thermal velocity dispersion is $\sim0.5$~\kms\ in both hubs, suggesting that the effects of turbulence are the same. 
While turbulence at larger scales could be responsible for the cloud fragmentation into filamentary structures, at smaller scales gravity might be more relevant. In fact, \citet{busquet2013} show that the fragmentation of filaments constituting the cloud complex IRDC G14.2 can be described by the fragmentation of an isothermal cylinder determined by thermal pressure with a (small) additional contribution from turbulent pressure. High angular resolution observations of molecular lines would provide information on the relative velocity between the different cores embedded in the hub, which would confirm the nature of the internal gas motions inside each core (Busquet \et\ in prep.).

Another important physical agent that presumably determines the
fragmentation level in a cloud core is the magnetic
field. Observationally, \citet{palau2013} investigate the effects of
the magnetic field in a sample of dense cores and find that the
low-fragmented regions are well reproduced in the magnetized core
case, while the highly fragmented regions are consistent with cores
where turbulence dominates over magnetic field as indicated by a
comparison with magnetohydrodynamic simulations
\citep{commercon2011}. Recent studies toward high-mass star-forming
regions also suggest that magnetic field plays a crucial role during
the collapse and fragmentation of massive clumps and the formation of
dense cores \citep{zhang2015,Li2015}. Regarding the IRDC G14.2, as
shown in \citet{busquet2013}, the network of parallel filaments traced
by the \nh\ dense gas could be the result of gravitational instability
of a thin gas layer threaded by magnetic fields. In particular,
magnetic fields seem to play an important role in the alignment of
filaments as revealed by the preliminary near-infrared polarization
measurements of a small part of the cloud. A follow up work of optical
and near-infrared polarization data, covering the entire group of
filaments, provides a measurement of the sky-projected field strength
in each hub (Santos \et\ in prep.). The inferred values are similar,
$\sim1$~mG in hub-N compared to 1.5~mG in hub-S, and the derived Alfv\'en Mach numbers are smaller than 1, suggesting that  during the fragmentation of the cloud core the magnetic field is an important ingredient that should not be ignored. Actually, Santos \et\ show that magnetic fields are tightly perpendicular to the dense filaments and hubs but also to the cloud as a whole. 
However, polarization properties could be different when tracing
different scales, \ie\ different parts of the cloud
\citep[\eg][]{alves2014}, and hence the magnetic field deep into the
dense hubs could be different from that inferred from the
near-infrared polarimetric data. This might have the consequence that
a stronger magnetic field pervades hub-N relative to hub-S resulting
hub-N undergoing lesser fragmentation, a scenario that could be tested with ALMA. 

Therefore, the physical properties investigated in this section,  \ie\ density structure, turbulence, and magnetic field, are remarkably similar in both hubs. We stress that we assumed that the magnetic field properties derived from optical and near-infrared data hold deep into the dense hubs, so any change in such properties, specially differences between the two hubs, could lead to different level of fragmentation in this region.

\subsection{Evolutionary stage and UV radiation feedback effects}

Another possible scenario that could explain the observed differences in the fragmentation level would be the effects of UV radiation from nearby high-mass stars and/or differences in the evolutionary stage. 

As shown in Fig.~\ref{g14-nh311}, hub-N is located near the infrared
source IRAS\,18153--1651, which has a luminosity of
$\sim1.1\times10^4$~\lo\ \citep{jaffe1982}. The VLA 6~cm continuum
emission reveal a cometary \hii\ region ionized by a B1 star with the
head of the cometary arch pointing toward hub-N (A. Sánchez-Monge
2012, private communication) that could compress and heat the gas. In
fact, \citet{busquet2013} find evidence for  local heating in the
western part of hub-N facing the \hii\ region, and the
position-velocity plot of the \nh\ dense gas along hub-N unveils an
inverted C-like structure consistent with an expanding
shell. Numerical simulations suggest that radiative heating from nearby high-mass protostars can strongly suppress fragmentation \citep[\eg][]{offner2009,hansen2012,myers2013}. Radiative feedback from this IRAS source could thus explain why hub-N is less fragmented than hub-S since an increase in temperature would increase the Jeans mass and consequently the Jeans number would be smaller. However, both temperature and Jeans mass are extremely similar in these two hubs, suggesting that radiative feedback does not seem to play a dominant role in the fragmentation process of hub-N, and further observations are needed to investigate
whether the UV radiation from the nearby IRAS source is affecting the properties of hub-N. Since molecular outflows are unknown in this region, we cannot account for their feedback effects. 


Finally, the evolutionary stage of these hubs could also lead to the differences observed in the fragmentation level. We searched infrared sources classified as candidate to YSO by \citet{povich2010} located within the SMA field of view. 
In Figs.~\ref{fsmacont-hubN} and \ref{fsmahubS-comb}, we show the position and the evolutionary stage (using different colors and symbols) of the YSOs identified in each hub. Hub-N contains 14 infrared sources whereas only 10 sources are associated with hub-S. 
The main difference arises in the number of infrared sources classified in the Stage 0/I (dominated by an infalling envelope): there are 10 YSOs in hub-N while hub-S contains only 6 YSOs in this stage of evolution. In both hubs, there are 3 YSOs in the more advanced evolutionary stage (Stage II, which have optically thick cicumstellar disk), and in each hub one source has been classified as ambiguous. Interestingly, the YSOs in hub-N appear to be more clustered around the bright millimeter condensation MM1a (see left panel in Fig.~\ref{fmmreg}). 

We report in Table~\ref{tquant} the ratio between the number of infrared sources without a millimeter counterpart and the number of millimeter sources within a region 0.3~pc of diameter. In this case, we do see differences between both hubs:  $N_{\rm{IR}}/N_{\rm{mm}}\simeq1.7$ in hub-N in front of 0.4 in hub-S. Thus, the relative number of infrared sources in front of the millimeter sources is larger in hub-N by a factor $\sim4$. 
Most of the objects associated with hub-N are more evolved YSOs than in hub-S, which still harbors deeply embedded objects in an earlier evolutionary stage. Therefore, the cluster of low-mass protostars detected in the infrared and associated with hub-N could effectively heat the gas and thus prevent fragmentation \citep{krumholz2006,krumholz2007}. 

Thus, the evolutionary stage of each hub could provide an explanation to the observed differences in the fragmentation level. Clearly, this scenario should be further investigated together with the information of the magnetic field at small scale.

\section{Conclusions \label{sconcl}}

We present 1.3~mm continuum observations carried out with the SMA toward hub-N and hub-S of the IRDC\,G14.225-0.506 complemented with single-dish continuum observations at 870 and 350~\mum\ conducted with the APEX and CSO telescopes. The main findings of this work are summarized as follows.

\begin{enumerate}

\item The two hubs presented in this work show different level of fragmentation in the millimeter range. 
Toward hub-N we find 4 millimeter condensations, while hub-S displays a higher degree of fragmentation, with 13 millimeter condensation detected within a field of view of 0.3~pc of diameter.  

\item We fitted simultaneously the radial intensity profiles at 870 and 350~\mum\ and the SED assuming spherical symmetry. In both hubs the radial density is well described by a Plummer-like function with a power-law index of 2.2, a central density of $\sim1.3\times10^{-18}$~g\,\cmt, and a radius of the flat inner region of 21000~AU, while temperature can be described as a power-law function with a power-law index of 0.34 and a dust emissivity index of 1.8. The differences in the fragmentation level cannot be explained by the density and temperature structure of these hubs. 

\item The CFE is 3\,\%\ and 13\,\%\ in hub-N and hub-S, respectively.  The derived masses and the observed number of fragments are consistent with pure thermal Jeans fragmentation, while adding non-thermal support implies that these hubs should not fragment. The level of fragmentation can thus be explained without invoking the presence of turbulent support.  

\item All the derived properties
such as the level of turbulence and magnetic field are remarkably similar in these hubs. In particular, the non-thermal turbulent contribution to the velocity dispersion is $\sim0.6$~\kms\ in both hubs, and despite internal motions are slightly supersonic they do not seem to dominate at scales of 0.15~pc. 
On the other hand, both the value of the magnetic field as well as its orientation with respect to the main hub axis are remarkably similar in both hubs. Differences in the polarization properties at smaller scales, not traced by our $H$-band polarization data, could provide a reasonable explanation to the observed differences in the fragmentation level: a stronger magnetic field would suppress fragmentation, as it would be the case of hub-N.

\item Evolutionary effects and possibly feedback effects from the UV radiation of a nearby massive B1 star
in hub-N could be an alternative explanation for the observed differences in the fragmentation level in these hubs.

\end{enumerate}

In summary, our analysis of the fragmentation level in the twin hubs of the IRDC G14.2 reveals that the fragmentation process at scales $\ga2000$~AU is more consistent with pure thermal Jeans fragmentation than with fragmentation including turbulent support, and that agents such as the magnetic field, the evolutionary stage, and radiative feedback might be crucial and need to be considered in future work.\\

\acknowledgments
\begin{small}
The authors are grateful to the anonymous referee for providing comments improving the clarity and quality of the paper.
G.B is grateful to Sergio Molinari and Davide Elia, from the Hi-GAL project, for providing the \textit{Herschel} Hi-GAL fluxes to build the SEDs. 
G.B. acknowledges the support of the Spanish Ministerio de Economia y Competitividad (MINECO) under the grant FPDI-2013-18204.
G.B., R.E., J.M.G., and G.A. are supported by the Spanish MICINN grants AYA2011-30228-C03 and AYA2014-57369-C3 
(co-funded with FEDER funds).  A.P. acknowledges financial support
from the UNAM-DGAPA-PAPIIT IA102815 grant, M\'exico. T.P. acknowledges financial support from the \emph{Deut\-sche For\-schungs\-ge\-mein\-schaft, DFG\/}  via the SPP (priority program) 1573 ("Physics of the ISM").

Facilities: SMA, APEX (LABOCA), CSO (SHARC~II)

\end{small}

\bibliographystyle{apj} 

\bibliography{g14_sma.bib}

\end{document}